\documentclass[acmsmall]{acmart}

\acmJournal{PACMPL}
\acmVolume{1}
\acmNumber{POPL}
\acmArticle{1}
\acmYear{2020}
\acmMonth{1}
\acmDOI{} 
\startPage{1}

\setcopyright{none}

\usepackage{booktabs}
\usepackage{shuffle}
\usepackage{stmaryrd}
\usepackage{subcaption}

\usepackage[boxed]{algorithm2e}
\usepackage{array}
\usepackage{amssymb}
\usepackage[bibliography=common,appendix=inline]{apxproof}
\usepackage{bussproofs}
\usepackage{caption}
\usepackage{centernot}
\usepackage{float}
\usepackage{longtable}
\usepackage{listings}
\usepackage{microtype}
\usepackage{multicol}
\usepackage{multirow}
\usepackage{pdflscape}
\usepackage{tikz}
\usepackage{tikz-qtree}
\usepackage{wrapfig}

\begin{document}

\title{Reductions for Safety Proofs (Extended Version)}

\author{Azadeh Farzan}
\affiliation{
  \institution{University of Toronto}            
}
\email{azadeh@cs.toronto.edu}          

\author{Anthony Vandikas}
\affiliation{
  \institution{University of Toronto}
}
\email{anthony.vandikas@mail.utoronto.ca}


\newcommand{\red}[3]{{#1}{\downarrow_{#2,#3}}}
\newcommand{\st}{\mathcal{S}t}
\newcommand{\sset}{sleep~}
\newcommand{\ord}{O}
\newcommand{\sred}{{S-reduction}}
\newcommand{\sRed}{{S-Reduction}}

\newcommand{\pow}[1]{{\mathcal{P}(#1)}}
\newcommand{\lang}[2][]{\mathcal{L}_{#1}(#2)}
\newcommand{\sem}[1]{\llbracket#1\rrbracket}
\newcommand{\floor}[1]{{\lfloor#1\rfloor}}
\newcommand{\ceil}[1]{{\lceil#1\rceil}}
\newcommand{\linear}[1]{\mathcal{L}in(#1)}

\newcommand{\tool}{{\sc Slacker}~}

\newcommand{\semc}{\sqsubseteq}

\def\@proofindent{\noindent}

\renewcommand{\state}{\mathcal{S}t}
\newcommand{\assert}{\mathcal{A}}
\newcommand{\stmt}{\Sigma}

\newcommand{\bool}{\mathbb{B}}
\newcommand{\true}{\top}
\newcommand{\false}{\bot}

\newcommand{\fst}{\textrm{fst}}
\newcommand{\snd}{\textrm{snd}}

\newcommand{\partition}[1]{\mathcal{P}art(#1)}

\newcommand{\action}[1]{$\mathtt{#1}$}

\newcommand{\sleep}{\mathrm{sleep}}
\newcommand{\ignore}{\mathrm{ignore}}
\newcommand{\reduce}{\mathrm{SRed}}
\newcommand{\creduce}{\mathrm{CRed}}
\newcommand{\inactive}{\mathrm{inactive}}

\newcommand\preceqdot{\mathrel{\ooalign{$\le$\cr
  \hidewidth\raise0.15ex\hbox{$\cdot\mkern0.3mu$}\cr}}}

\def\marrow{{\marginpar[\hfill$\longrightarrow$]{$\longleftarrow$}}}
\def\azadeh#1{}
\def\anthony#1{}

\newtheoremrep{theorem}{Theorem}[section]
\newtheoremrep{lemma}[theorem]{Lemma}
\newtheoremrep{proposition}[theorem]{Proposition}

\begin{abstract}
{\em Program reductions} are used widely to simplify reasoning about the correctness of concurrent and distributed programs. In this paper, we propose a general approach to proof simplification of concurrent programs based on exploring {\em generic} classes of reductions. We introduce two classes of sound program reductions, study their theoretical properties, show how they can be effectively used in algorithmic verification, and demonstrate that they are very effective in producing proofs of a diverse class of programs without targeting specific syntactic properties of these programs. The most novel contribution of this paper is the introduction of the concept of {\em context} in the definition of program reductions. We demonstrate how {\em commutativity} of program steps in some program contexts can be used to define a generic class of sound reductions which can be used to automatically produce proofs for programs  whose complete Floyd-Hoare style proofs are theoretically beyond the reach of automated verification technology of today.
\end{abstract}

%
%

\maketitle

\section{Introduction}\label{sec:intro}
A {\em reduction} of a program is generally another program, with a subset of the behaviours of the original program, that faithfully represents it. Program reductions have been studied extensively \cite{Lipton75,ElmasQT09,civl,Desai14,psynch,Genest07} in the context of simplifying reasoning about concurrent and distributed programs. The earliest and perhaps most well-known approach to reduction is due to Lipton \cite{Lipton75} who proposed to simplify concurrent program proofs by inferring large atomic blocks of code (when possible) in order to reap the benefits of sound sequential reasoning inside these blocks. The inference of the large atomic blocks is carried out based on commutativity specifications of individual program statements. In the past 40 years, Lipton's work has inspired many reduction schemes for concurrent program analysis \cite{FlanaganFQ05,FlanaganQ03} and verification \cite{ElmasQT09,civl}. In a different context, commutativity specification of program statements have been used for an entirely different type of reduction. There, the aim is to reduce the sizes of the communication buffers used in message-passing programs. The equivalent program with smallest buffer sizes can be viewed as an {\em almost synchronous} variation of the original asynchronous program. The key insight is that the proof of correctness for the synchronous program is {\em simpler}; for program with bounded buffers the proof need not include complex invariants such as those that universally quantify over unbounded buffer contents.


The two groups of reduction approaches strive for seemingly contradictory targets.
Lipton's approach opts for reductions in which the threads try not to {\em yield} for as long as possible, while synchronous reductions would force a yield right after each {\em send} operation in order to execute its matching {\em receive}. The former seems appropriate for shared memory concurrent programs and the latter for message-passing concurrent and distributed programs. This sparks several interesting questions: is this truly a rigid dichotomy? Can shared memory concurrent programs benefit from certain types of synchronous reductions where arbitrary program statements (other than just sends and receives on channels) are synchronized? What sort of reductions can deal with message-passing concurrent programs where reasoning has to be extended to the part of the program that manipulates the data? Can we not commit to a particular reduction scheme in advance and let the verifier pick the ideal reduction for the input program, depending on what is required for the the specific combination of the program and the property? In this paper, we provide some initial answers to these questions.

We propose an automated verification approach that combines the search for a proof with the search for a sound reduction of the program. The high level idea is to give a chance to automated verification to succeed by finding a correctness proof for a reduction of the program, where it would fail otherwise if it attempted to prove the original program correct. The key distinction with regards to most of the relevant literature is that instead of fixing a particular reduction in advance, we propose to let a new  automated verification algorithm search for an ideal reduction within a generic universe of (infinitely many) sound reductions. The simple insight is that committing to the wrong reduction in advance, for example attempting to infer large atomic blocks for a distributed message-passing program, could set one up for failure.
Our target programs are principally those where {\em proof simplification is the difference between the existence and nonexistence of a safety proof} within a fixed (decidable) language of assertions commonly used in automated verification. Without simplification, the proof involves complicated invariants with elements such as quantification over arrays and buffers or non-linear arithmetic for data variables which are currently the Achilles heel of automated verification techniques. Therefore, the main accomplishment of our methodology is to leverage the proof simplification power of reductions to expand the reach of automated verification to instances that are theoretically out of its scope.

Our refinement loop maintains a proof candidate at each round, and checks if there exists a reduction of the input program that is proved correct by this proof. To be able to implement this subsumption test algorithmically, one needs an effective way of representing the set of all program reductions. We introduce two novel classes of (infinitely many) program reductions and use finite state (tree) automata, with nice algorithmic properties, to represent each class. The first class is inspired by semi-trace monoids \cite{diekert1995book} defined by a semi-commutativity relation between program statements.  Unfortunately, checking whether a proof subsumes a reduction of the program according to such a semi-trace monoid is in general undecidable (more on this in Section \ref{sec:semi}). Therefore, the contribution of this paper critical to algorithmic verification is devising a subclass, which we call {\em S-reductions} with a decidable subsumption check.

The most significant contribution of this paper is the second proposed class of reductions, namely {\em contextual reductions}. For this class, the commutativity properties of the program statements depend on the context from which the corresponding statements are executed. Two statements may commute in one {\em context} and not in another. Contexts have been exploited in special cases for proofs before. For example, in message-passing programs, a {\em receive} can be commuted to the left of a {\em send} operation that is not its matching {\em send}, determined by by context.

To the best of our knowledge, general contexts have never been exploited for program proofs before, and certainly not for automated verification.

Inspired by a language-theoretic notion of context from {\em generalized Mazurkiewicz traces}, we define a set of (infinitely many) contextual reductions that is recognized by finite state automata. The elegance of this definition is that it does not commit to a particular contextual commutativity relation in advance. The automaton models a universe of reductions based on a universe of contextual commutativity specifications.
Our proposed algorithm then decides if there exists a sound contextual commutativity specification in this universe, which induces a sound contextual reduction of the program that is covered by a current valid proof candidate.
Therefore, beyond making progress in proof construction, the refinement loop also infers assertions that substantiate the soundness of a larger contextual commutativity relation through refinement. This goes against the classical approach to automated program verification where one first chooses a (static) mostly non-contextual commutativity specification, {\em then} a reduction induced by the chosen specification, and {\em finally then} tries to search for a proof for the given reduction. Our proposed refinement loop performs all these three searches simultaneously.
In summary, the following are the contributions of this paper:
\begin{itemize}
\item We introduce a class of semi-commutative reductions and present the theoretical properties of this class that make it a good candidate for algorithmic proof simplification (Section \ref{sec:semi}).
\item We introduce a novel class of contextually commutative reductions essential to proof simplification for both shared-memory and message-passing concurrent programs, and theoretically argue why they are suitable for algorithmic use (Section \ref{sec:context}).
\item We present a counterexample-guided refinement loop for verification which can incorporate the above classes of reductions.
This algorithm effectively performs a search for a triple consisting of {\em a contextual commutativity relation, a program reduction induced by it, and a proof of correctness for the reduction}.
We discuss the soundness, completeness, and convergence conditions for the algorithm. Moreover, we present two interesting insights that accommodate the development of a novel algorithm for proof checking with an improved time complexity upper bound. (Sections \ref{sec:algorithm} and \ref{sec:optimization}).
\item We provide an in-depth comparison of the reductions presented in this paper and the two most well-known reduction schemes from the concurrent program verification literature (Section \ref{sec:relations}): (1) Lipton's reduction for the inference of large atomic blocks, and (2)  reductions based on the idea of existential boundedness \cite{Genest07} which use commutativity-based transformations to reduce message buffer sizes to simplify proofs of message-passing concurrent/distributed programs.
\item Our approach is implemented in a tool, called \tool. Using a rich set of benchmarks,  mostly with required invariants beyond the reach of previous automated verification tools, we demonstrate how the technique is effective in producing  (automatically generated) proofs for these benchmarks (Section \ref{sec:exp}).
\end{itemize}

\section{Motivating Examples}\label{sec:example}

We start by motivating the two classes of reductions proposed in this paper through two examples.  In our first example, proof simplification is not essential, in that a proof for the program exists and can be discovered using a standard verification algorithm \cite{HeizmannHP09}. By using the class of semi-commutative reductions in this paper, however, one can produce a simpler proof (about half the number of distinct assertions in the proof) in less than one third of the time. We then
\begin{wrapfigure}{r}{0.33\textwidth}\vspace{-15pt}
\begin{center}
\includegraphics[scale=0.27]{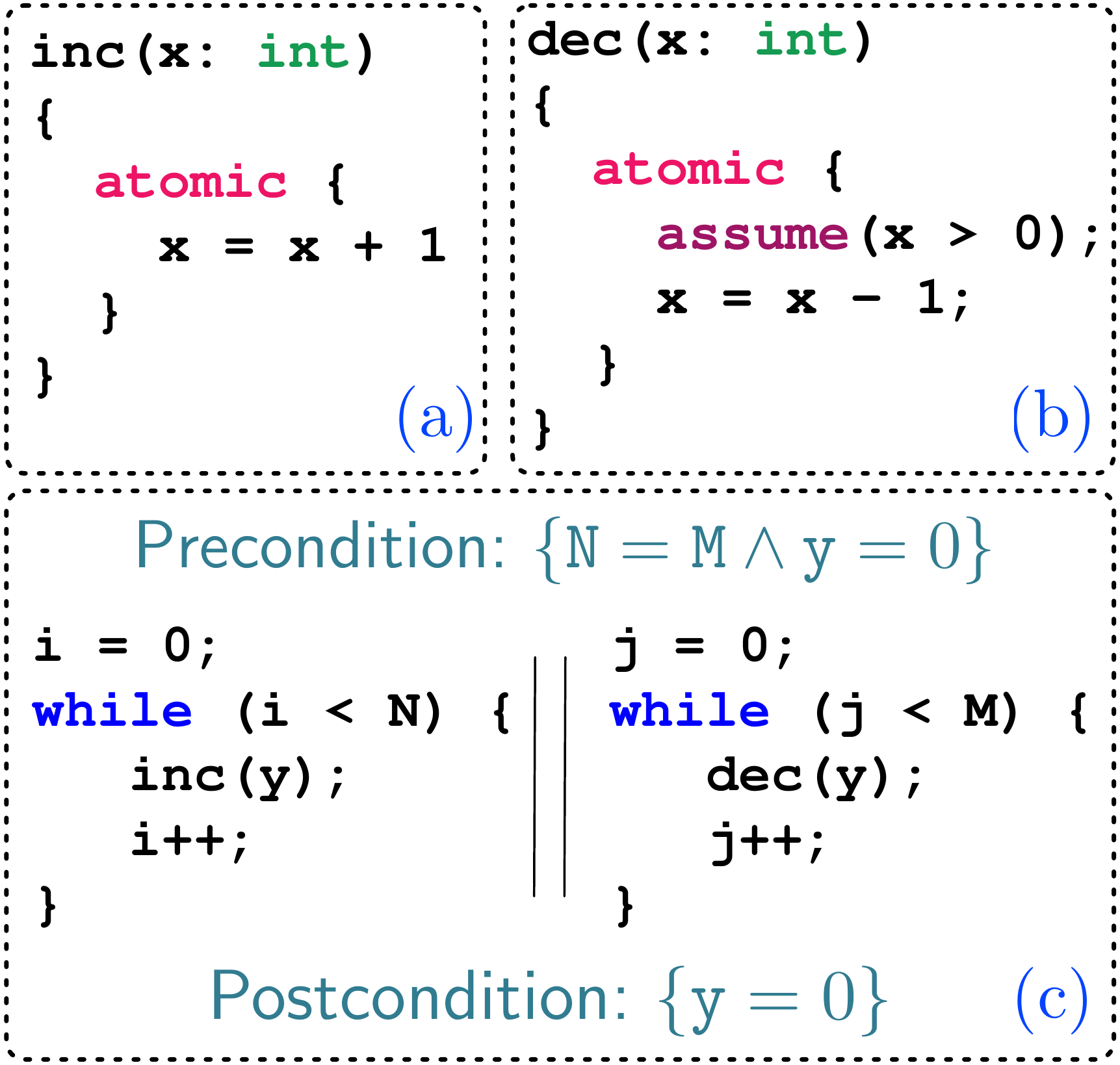}\vspace{-10pt}
\caption{ {\footnotesize Semi-commutativity example.}} \label{fig:me1}\vspace{-10pt}
\end{center}\vspace{-5pt}
\end{wrapfigure}
make a small modification to the code to get our second example, for which a proof for the whole
program does not exist in the decidable assertion language of linear integer arithmetic (LIA). Moreover, even though the program does admit a semi-commutative reduction, a proof does not exist for that reduction either. This will motivate our {\em contextual reduction} class, which includes a sound reduction of the program with a simpler proof that is quickly discovered by \tool.

Consider the simple methods {\tt inc()} and {\tt dec()} defined in Figure \ref{fig:me1}(a,b), and a simple concurrent program using them listed in Figure \ref{fig:me1}(c), along with its corresponding pre/post-conditions. Note that {\tt dec()} is a blocking statement and therefore not all program runs terminate.  For a safety proof, it suffices to show that those that do satisfy the given pre/post-condition. It is straightforward to see that a full Floyd-Hoare style proof of this program exists in the decidable logical language of linear integer arithmetic.

Observe that {\tt inc()} soundly semi-commutes with {\tt dec()} in the sense that it is sound to swap a {\tt dec()} statement to the right of a following {\tt inc()} statement,  without changing program behaviour. The inverse, however, is not true. Swapping a decrement to the left of an increment {\em may} make it block in some program runs where it was not blocking in its original position. Adding this to the fact that {\tt i} and {\tt j} are thread-local and therefore all statements referencing them commute (against the statements of the other thread), indicates that a full proof for the program is not strictly necessary. It is sufficient to provide a proof for a subset of the program runs that soundly represent the program, and this subset may admit a strictly simpler proof. The discovery of this simple proof is the goal of the methodology presented in this paper.

\begin{wrapfigure}{r}{0.18\textwidth}\vspace{-10pt}
\begin{center}
\includegraphics[scale=0.28]{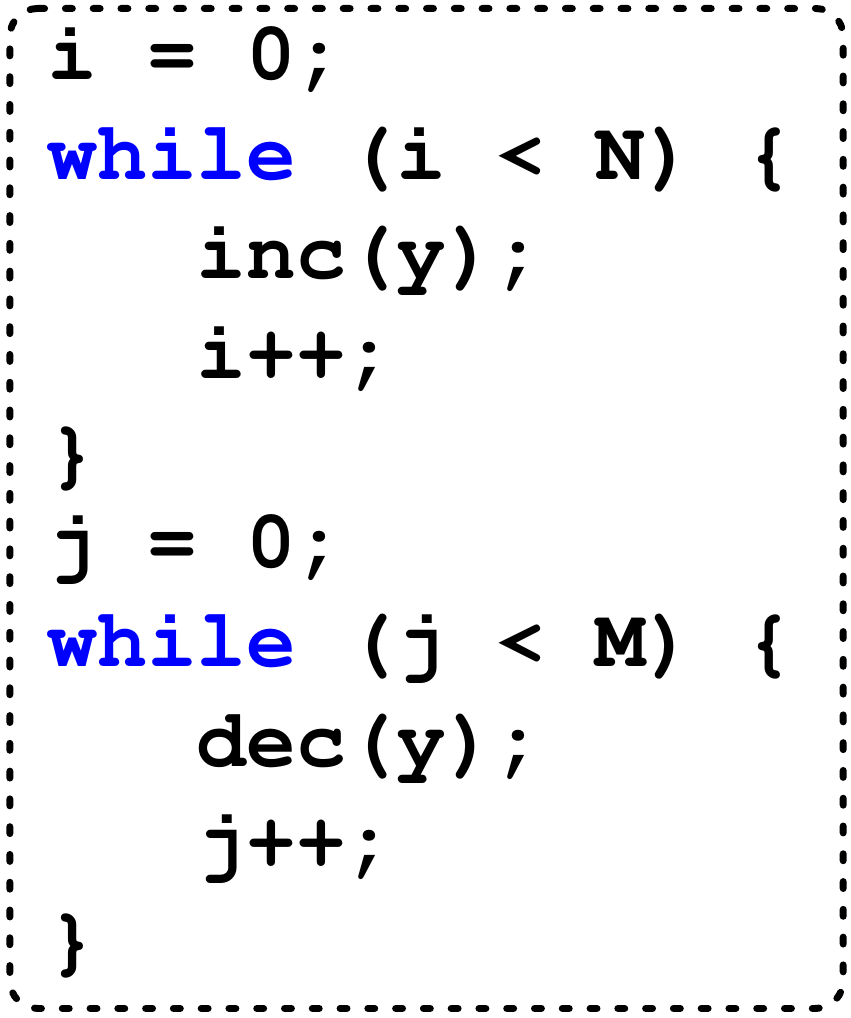}\vspace{-10pt}
\caption{A reduction.} \label{fig:seq}\vspace{-5pt}
\end{center}\vspace{-10pt}
\end{wrapfigure}

The sequential program illustrated in Figure \ref{fig:seq} is a sound {\em reduction} of the program in Figure \ref{fig:me1}. In this reduction, {\em all} decrements are postponed to the end using the semi-commutativity of {\tt dec()} and
{\tt inc()}. The full commutativity of the rest of the actions is then used to bring relevant steps of each
thread together.  Our proposed set of {\em S-reductions}
 includes this sequential program as well as (infinitely) many other reductions that are equivalent to the program up to the aforementioned (semi-) commutativity properties of the statements. Our proposed algorithm attempts to verify at least one member of the entire set in a refinement loop. Our tool, \tool, discovers a simpler proof (about half the number of distinct assertions in the proof) in about a third of the time of the original (without reductions). Note that the reduction, for which \tool discovers a proof may not match the one in Figure \ref{fig:seq} precisely.

The reader familiar with Lipton's reductions \cite{Lipton75} and the concept of left/right movers would be curious about the connection between this transformation and Lipton's atomic blocks reductions. Note that  {\tt dec()} is a right-mover, and respectively, {\tt inc()} is a left mover, and every other statement is both \footnote{In fact, since Lipton's original definition in \cite{Lipton75} is quantified over all reachable program contexts, {\tt inc()} and {\tt dec()} would be both-movers according to his original definition. But, folklore usage of his technique, which quantifies over all contexts (reachable or not), would declare them only left and right mover respectively. }.
Therefore, one can soundly declare each thread as one atomic block and end up with a reduced program that runs these two atomic blocks in parallel. This program has additional behaviours compared to the (sequentialized) reduction of Figure \ref{fig:seq}. The difference is not substantial in this case. Next, we will look at a slight modification of this program which would
\begin{wrapfigure}{r}{0.4\textwidth}\vspace{-15pt}
\begin{center}
\includegraphics[scale=0.28]{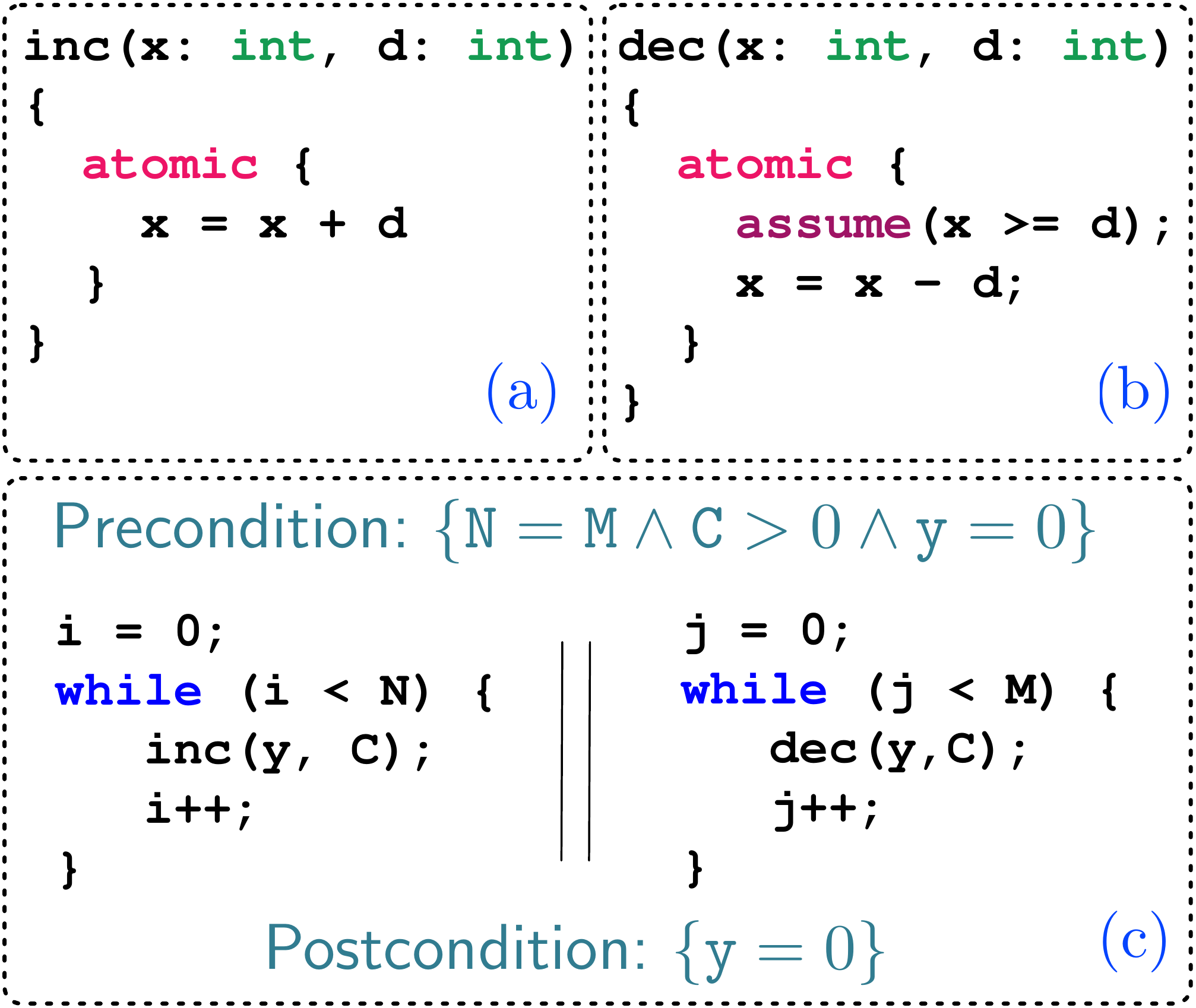}\vspace{-10pt}
\caption{Contextual commutativity example.} \label{fig:me2}\vspace{-10pt}
\end{center}\vspace{-5pt}
\end{wrapfigure}
render both Lipton style reductions and our S-reductions entirely useless for proof simplification.

Consider the modified code illustrated in Figure \ref{fig:me2}. The methods {\tt inc()} and {\tt dec()} operate as before, but now take an extra parameter determining the increment/decrement {\em delta}. The program uses a global (uninitialized) positive constant {\tt C} as this delta, and otherwise operates as before. Note that this program admits the same sequential reduction in the style of Figure \ref{fig:seq}, since the new {\tt inc()} and {\tt dec()} methods satisfy the same (semi-) commutative properties as in the previous example. The problem is, however, that this sequential reduction does not admit a proof in the decidable LIA fragment. The proof needs to establish at the end of the first loop that $\mathtt{y} = \mathtt{N} \times \mathtt{C}$, so that by the end of the second loop, {\tt y} can be proved to go back to zero. This requires a non-linear loop invariant $\mathtt{y} = \mathtt{i} \times \mathtt{C}$ for the first loop. Lipton's reductions are also not effective for the same reason. Luckily, there is another reduction of this program that does admit a proof in the LIA fragment.

Any program trace that has a different number of increments and decrements is infeasible, and this can be reflected in the proof by simple invariants relating {\tt i}, {\tt j}, {\tt M}, and {\tt N}. All feasible program traces will have an equal number of increments and decrements.  To avoid having to use multiplicative assertions like $\mathtt{y} = \mathtt{N} \times \mathtt{C}$, of all the equivalent feasible interleavings of the program, the one in which increments and decrements appear in alternate order is the preferred one:
\begin{quote}
{\tt inc(y,C)} \dots {\tt dec(y,C)} \dots {\tt inc(y,C)} \dots {\tt dec(y,C)} \dots.
\end{quote}
For these interleavings, the invariants need to only capture the fact that {\tt y} goes up by {\tt C} and then comes back to zero when the matching decrement happens. The reduction that only includes these interleavings is in some sense  {\em the opposite} of the sequential reduction of Figure \ref{fig:seq}. In the sequential one, an entire thread is executed as an atomic block, while in this one, threads are forced to yield after each increment to let the matching decrement execute. It is interesting how a small change in the program can have a big impact on the appropriate reduction for proving it correct.

Let us now argue why the suggested reduction is sound. The key is  the concept of {\em contextual commutativity}. Note that {\tt inc(y)} and {\tt dec(y)} fully commute if $\mathtt{y} \ge \mathtt{C}$. Only for values of $\mathtt{y} < \mathtt{C}$, do they semi-commute (as discussed above). Under this contextual commutativity relation, one can show that the interleaving proposed above is equivalent to all other interleavings of the program.  The high level argument is: we already know that all decrements can be postponed to the end, due to the (non-contextual) semi-commutativity relation. Therefore, every interleaving is equivalent to one with all the decrements appearing at the end. Starting from that interleaving, the decrements can be pulled forward one by one to appear next to a (matching) increment, because we know $\mathtt{y} \ge \mathtt{C}$ is true before each decrement (that has a matching increment in the prefix of the run). Note that this last step cannot be performed under the static semi-commutativity assumption. A decrement does not commute to the left of an increment.

The main observation is that {\em contexts matter}. At the beginning, before any increments or decrements have been executed, the two operations do not commute (when $\mathtt{y} = 0$). Once an increment is executed, then $\mathtt{y} \ge \mathtt{C}$ is established and then the operations commute.

Our proposed set of contextual reductions, called {\em C-reductions}, includes this preferred reduction and (infinitely) many more equivalent ones. In a refinement loop, our algorithm infers such contextual commutativity information, and uses it to discover a sound contextual reduction of the program that can be proved correct. The proof is in the pudding: the algorithm decides which reductions are sound and among those which can be proved correct by actually producing proofs of soundness of reductions and correctness of at least one specific reduction.
\tool can discover a proof for the program in Figure \ref{fig:me2} in a few seconds.


\section{Background}\label{sec:background}

\subsection{Programs and Proofs}


\subsubsection*{\bfseries Programs as Regular Languages}\label{sec:ptrace}

$\state$ denotes the (possibly infinite) set of \emph{program states}. For example, we have $\state = \mathbb{Z} \times \mathbb{Z}$ for a program with two integer variables. Let $\assert \subseteq \pow{\state}$ be a (possibly infinite) set of \emph{assertions}.  $\stmt$ denotes a finite alphabet of program \emph{statements}. For multithreaded programs, statements are annotated with thread identifiers to distinguish the same statement of different threads. We assume a bounded number of threads.

We refer to a finite string of statements as a (program) \emph{trace}. For each statement $a \in \stmt$, we associate a \emph{semantics} $\sem{a} \subseteq \state \times \state$ and extend $\sem{-}$ to traces via (relation) composition. A trace $\tau \in \stmt^*$ is said to be \emph{infeasible} if $\sem{\tau}(\state) = \emptyset$, where $\sem{\tau}(\state)$ denotes the image of $\sem{\tau}$ under $\state$. Note that the set of program traces is a superset of the set of concrete program executions (i.e. feasible program traces).

Without loss of generality, we define a \emph{program} as a language of traces. The semantics of a program $P$ is simply the union of the semantics of its traces $\sem{P} = \bigcup_{x \in P} \sem{x}$. Concretely, one may obtain the language of program traces by interpreting the edge-labelled control-flow graph of the program as a deterministic finite automaton (DFA): each location in the control flow graph is a DFA state, and each edge in the control flow graph is a DFA transition. The control flow graph entry location is the initial state of the DFA and all its exit locations are the DFA final states. We do not define programs to necessarily be \emph{regular} languages, but we do require our input programs to be regular and many important results require this.
\anthony{I think we may be inconsistent on whether we assume programs to be regular or not, especially since our reductions include non-regular languages. TODO: check sanity.}
\subsubsection*{\bfseries Program Safety}\label{sec:safety}

In the context of this paper, a program $P$ is \emph{safe} if all traces of $P$ are infeasible, i.e. $\sem{P}(\state) = \emptyset$.
Standard partial correctness specifications can be represented as safety via a simple encoding. Given a precondition $\phi$ and a postcondition $\psi$, the validity of the Hoare-triple $\{\phi\}P\{\psi\}$ is equivalent to the safety of $[\phi] \cdot P \cdot [\neg\psi]$, where $[]$ is a standard assume statement (or the singleton language containing it), and $\cdot$ is language concatenation.

A \emph{proof} is defined based on a finite set of assertions $\Pi \subseteq \assert$ that includes $\mathit{true}$ and $\mathit{false}$. One can associate a regular language to each set of assertions $\Pi$ by defining the NFA $\Pi_{NFA} = (\Pi, \stmt, \delta_\Pi, \mathit{true}, \{\mathit{false}\})$ where
\[\delta_\Pi(\phi_{pre}, a) = \{ \phi_{post} \mid \sem{a}(\phi_{pre}) \subseteq \phi_{post} \}.\]
We refer to $\lang{\Pi_{NFA}}$, abbreviated as $\lang{\Pi}$, as a {\em proof}. Intuitively, $\lang{\Pi}$ consists of traces that can be proven infeasible using only assertions in $\Pi$. The following proof rule is therefore sound \cite{HeizmannHP09,FarzanKP13,FarzanKP15}:
\begin{equation}\label{rule:safety}
  \AxiomC{$\exists \Pi \subseteq \assert \ldotp P \subseteq \lang{\Pi}$}
  \UnaryInfC{$P$ is safe}
  \DisplayProof
  \tag{\textsc{Safe}}
\end{equation}

When $P \subseteq \lang{\Pi}$, we say that $\lang{\Pi}$ is a proof for $P$. A proof does not uniquely belong to any particular program; a single language $\lang{\Pi}$ may prove many programs correct. When both $P$ and $\lang{\Pi}$ are regular, this check is decidable and polynomial on the sizes of their corresponding DFAs.

\subsection{Reductions}
\label{sec:reduction}

A safe program may not admit a safety proof in a given language of assertions, or it may admit one but the proof may be prohibitively complex. This has inspired the notion of {\em program reductions}. The reduction of a program $P$ is a simpler program $P'$  that may be soundly proved  safe in place of the original program $P$. Below is a very general definition of program reductions.

\begin{definition}[semantic reduction] \label{def:sr}
  If for programs $P$ and $P'$, $P'$ is safe implies that $P$ is safe, then $P'$ is a \emph{semantic reduction} of $P$ (written $P' \preceq P$).
\end{definition}

The definition immediately gives rise to the following sound proof rule for proving safety: 
\begin{equation}\label{rule:safety+reductions-bad}
  \AxiomC{$\exists P' \preceq P, \Pi \subseteq \assert \ldotp P' \subseteq \lang{\Pi}$}
  \UnaryInfC{$P$ is safe}
  \DisplayProof
  \tag{\textsc{SafeRed}}
\end{equation}

A program is safe if and only if $\emptyset$ is a valid reduction of the program, which means discovering a semantic reduction and proving safety are mutually reducible to each other. Therefore, verifying the existence of a {\em semantic reduction} is in general {\em undecidable}.
Therefore, a very particular choice of reduction is often used \cite{Desai14,psynch,Lipton75}.

There are instances in the literature \cite{Lipton75, cav19} where a restricted class of reductions have been used instead. For example, Lipton's reductions are technically a family of choices of atomic blocks based on left/write-movers in the program.
If one restricts the set of possible reductions from all reductions (given in Definition \ref{def:sr}) to a proper subset which more amenable to algorithmic checking, then the rule becomes more amenable to automation.
Fixing a set $\mathcal{R}$ of (semantic) reductions will change the rule to: 
\begin{equation}\label{rule:safety+reductions}
  \AxiomC{$\exists P' \in \mathcal{R} \ldotp P' \subseteq \lang{\Pi}$}
  \AxiomC{$\forall P' \in \mathcal{R} \ldotp P' \preceq P$}
  \BinaryInfC{$P$ is safe}
  \DisplayProof
  \tag{\textsc{SafeRed2}}
\end{equation}

In \cite{cav19} one candidate for $\mathcal{R}$ was presented in the form of a set of syntactic reductions which are called {\em sleep set} reductions. In this paper, we take a major step in defining a far more general (yet decidable) set of semantic reductions as a candidate for $\mathcal{R}$.

\subsection{Tree Automata for Classes of Languages}\label{sec:blta}
It is possible to automate the checking of the first premise of the rule \ref{rule:safety+reductions} through automata theoretic techniques.

\begin{wrapfigure}{r}{0.3\textwidth}\vspace{-15pt}
\begin{center}
\includegraphics[scale=0.26]{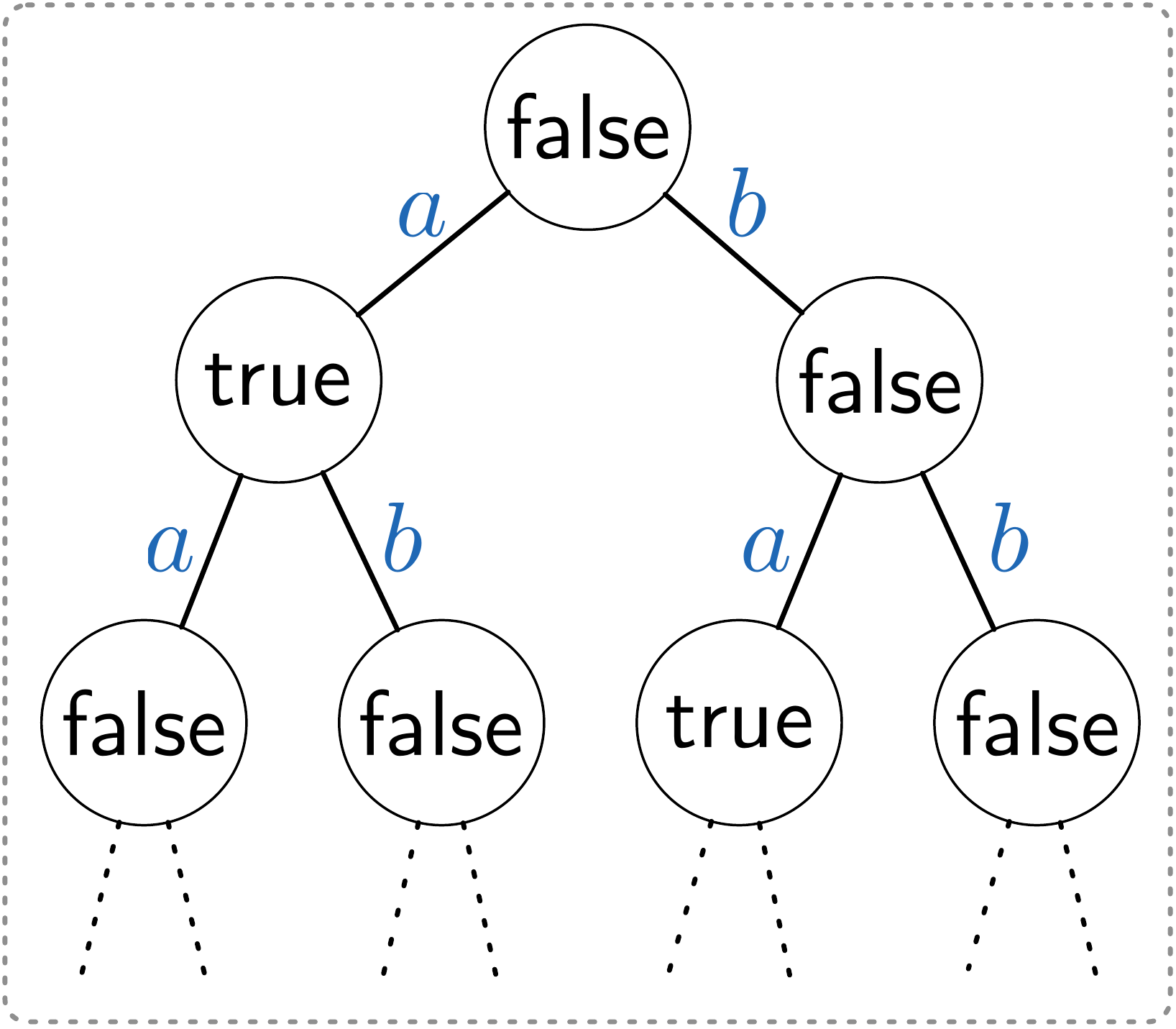}\vspace{-10pt}
\caption{Infinite tree representing the language $\{a,ba\}$. \label{fig:sse}}
\end{center}\vspace{-10pt}
\end{wrapfigure}
An infinite tree can encode a (potentially infinite) language of finite words. Consider the tree on the right where nodes are labeled with booleans and arcs are labeled with alphabet letters. A word belongs to the language represented by such a tree if it labels a path from the root to a {\em true} labeled node of the tree. A set of languages can then be encoded as a set of infinite trees. Certain sets of infinite trees are recognized by (finite state) automata over infinite trees. Looping Tree Automata (LTAs) are a subclass of B\"{u}chi Tree Automata where all states are accept states \cite{BaaderT01}. The class of Looping Tree Automata is closed under intersection and union, and checking emptiness of LTAs is decidable. Unlike B\"{u}chi Tree Automata, emptiness can be decided in linear time \cite{BaaderT01}.

\begin{definition}
A Looping Tree Automaton (LTA) over $|\stmt|$-ary, $\bool$-labelled trees is a tuple $M = (Q, \stmt, \Delta, q_0)$ where $Q$ is a finite set of states, $\Delta \subseteq Q \times \bool \times (\stmt \to Q)$ is the transition relation, and $q_0$ is the initial state.
\end{definition}

Formally, $M$'s execution over a tree $L$ is characterized by a \emph{run} $\delta^* : \stmt^* \to Q$ where $\delta^*(\epsilon) = q_0$ and $(\delta^*(x), x \in L, \lambda a \ldotp \delta^*(xa)) \in \Delta$ for all $x \in \stmt^*$. The set of languages accepted by $M$ is then defined as
  $\lang{M} = \{ L \mid \exists \delta^* \ldotp \text{$\delta^*$ is a run of $M$ on $L$} \}$.

\begin{theorem}[from \cite{cav19}]\label{thm:pc-decidable}
  Given an LTA $M$ and a regular language $L$, it is decidable whether $
    \exists P \in \lang{M} \ldotp P \subseteq L$.
\end{theorem}
Note that Theorem \ref{thm:pc-decidable} is effectively providing an automation recipe for the proof rule \ref{rule:safety+reductions}. In \cite{cav19}, a construction was given for a Looping Tree Automaton (LTA) that recognizes a specific family of reductions, called  {\em sleep-set reductions} of an input program $P$, which were shown to be useful in proving hypersafety properties of programs. In this paper, we will provide two extensions: a family of {\em static semi-commutative} reductions and a family of {\em contextual} reductions which are specifically useful for simplification of concurrent program proofs.


\section{Semi-commutative Reductions}\label{sec:semi}
\begin{toappendix}
  \label{app:semi}
\end{toappendix}
We introduce a class of reductions inspired by semi-commutative {\em Mazurkiewicz} traces (aka semi-trace monoids) and Lipton's \cite{Lipton75} left/right-movers.

Let $I \subseteq \stmt \times \stmt$ be an irreflexive (but not necessarily symmetric) \emph{semi-independence relation}.
Let $\semc_I$ be the smallest preorder satisfying $\sigma ab \rho \semc_I \sigma ba \rho$ for all $\sigma, \rho \in \stmt^*$ and $(a, b) \in I$. The upwards and downwards closures of a language $L \subseteq \stmt^*$ with respect to $\semc_I$ are respectively denoted by $\ceil{L}_I$ and $\floor{L}_I$ and defined as:
\begin{align*}
\ceil{L}_I &= \{u \mid \exists v \in L \ldotp v \semc_I u \} &\floor{L}_I = \{ u \mid \exists v \in L \ldotp u \semc_I v \}
\end{align*}
A language $L$ is \emph{upwards-closed} (resp. \emph{downwards-closed}) with respect to $\semc_I$ if $L = \ceil{L}_I$ (resp. $L = \floor{L}_I$).


If $I$ is a symmetric relation, then $\semc_I$ becomes an equivalence relation and its equivalence classes are known as \emph{Mazurkiewicz traces} \cite{DiekertM97}. As is the case with Mazurkiewicz traces, relation $I$ is of interest in program verification when it is \emph{sound}, i.e. $\sem{ab} \subseteq \sem{ba}$ for all $(a, b) \in I$.

\begin{definition}[semi-commutative reduction]\label{def:semc}
  A program $P'$ is a semi-commutative reduction of a program $P$, denoted by $P' \preceq_I P$, if $P \subseteq \floor{P'}_I$.
\end{definition}

Intuitively, in the reduction $P'$, it is safe to remove smaller traces (with respect to $\semc_I$) in favour of larger ones. $I$ is {\em sound} if $\sigma \semc_I \rho \implies \sem{\sigma} \subseteq \sem{\rho}$. Sound relations define sound reductions for safety verification. Formally:
\begin{lemmarep}\label{lem:semi-reduction}
  If $I$ is  a sound semi-independence relation and $P' \preceq_I P$ then $P' \preceq P$.
\end{lemmarep}
\begin{proof}
  Since $I$ is sound, it follows that $\sigma \sqsubseteq_I \tau$ implies $\sem{\sigma} \subseteq \sem{\tau}$ for any $\sigma, \tau \in \stmt^*$. Then for any $a, b \in \state$ we have
  \begin{align*}
    (a, b) \in \sem{P}
    &\implies \exists \sigma \in P \ldotp (a, b) \in \sem{\sigma} \\
    &\implies \exists \sigma \in \floor{P'}_I \ldotp (a, b) \in \sem{\sigma} \\
    &\implies \exists \sigma \ldotp \exists \tau \in P' \ldotp \sigma \sqsubseteq_I \tau \land (a, b) \in \sem{\sigma} \\
    &\implies \exists \sigma \ldotp \exists \tau \in P' \ldotp \sigma \sqsubseteq_I \tau \land (a, b) \in \sem{\tau} \\
    &\implies \exists \tau \in P' \ldotp (a, b) \in \sem{\tau} \\
    &\implies (a, b) \in \sem{P'}
  \end{align*}
  so $\sem{P} \subseteq \sem{P'}$ and therefore $P' \preceq P$.
\end{proof}

\begin{example}
Recall the example from Section \ref{sec:example} illustrated in Figure \ref{fig:me1}.
{\tt inc()} semi-commutes with {\tt dec()} in the sense that it would be sound to have $(\mbox{\tt dec()},\mbox{\tt inc()}) \in I$. But, the inverse is not true: $(\mbox{\tt inc()},\mbox{\tt dec()}) \not \in I$. The sequential program of Figure \ref{fig:seq} is a sound semi-commutative reduction of the program in Figure \ref{fig:me1}.
\end{example}

Ideally, the set of all (sound) semi-commutative reductions of a program would replace $\mathcal{R}$ in the premise of the rule \ref{rule:safety+reductions}. Unfortunately, this is not possible. It has already been argued in \cite{cav19} that the premise check $\exists P' \in \mathcal{R}. P' \subseteq \lang{\Pi}$ is undecidable for an arbitrary $\Pi$ for the special case where $I$ is symmetric. Considering our scenario is strictly more general, the undecidability result follows straightforwardly. Fortunately, there exists a suitable approximation of the set of {\em semi-commutative} reductions that can be used as a candidate for $\mathcal{R}$ rendering the premise decidable.

\subsection{A Representable Class of Semi-Commutative Reductions}\label{sec:repsem}

Recall from Section \ref{sec:blta} that a language can be represented by an infinite labelled tree, where the arcs are labelled with program statements. To reduce a language in a constructive way (in contrast to Definition \ref{def:semc}), one can prune this infinite tree in a style inspired by partial order reduction \cite{Godefroid96}. Pruning the tree is equivalent to removing words from the language, which defines a reduction.

\begin{figure}[ht]
\begin{center}\vspace{-10pt}
\includegraphics[scale=0.21]{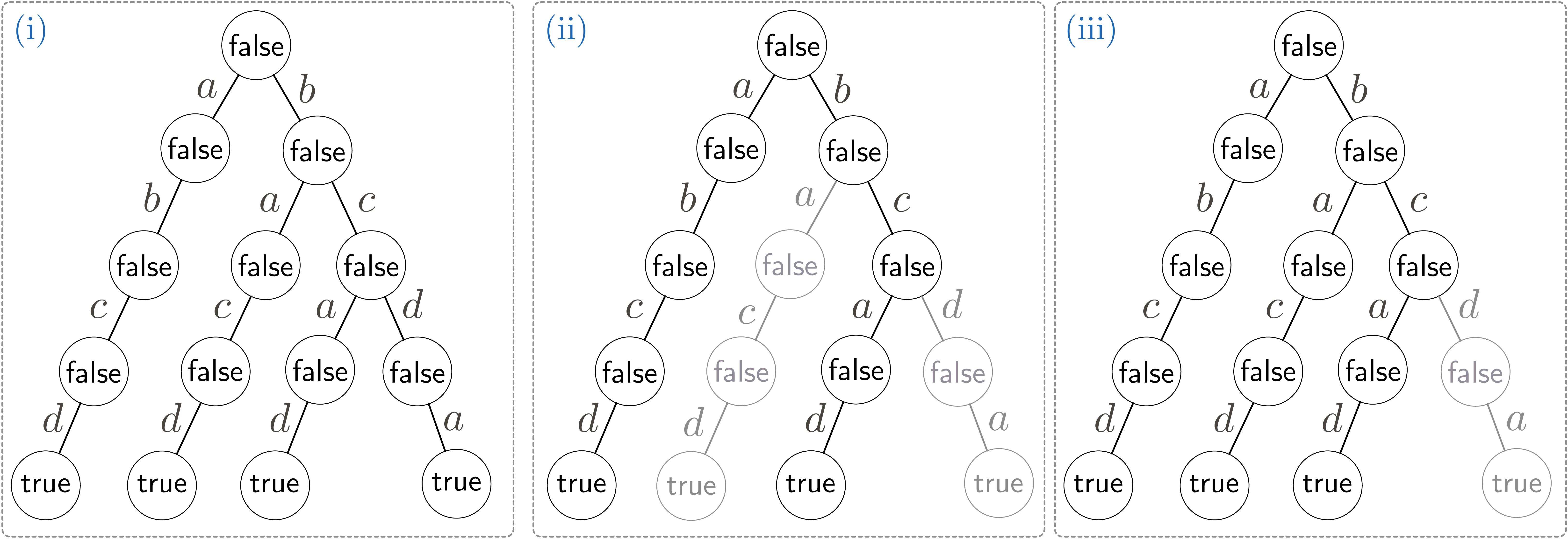}\vspace{-10pt}
\caption{An example illustrating reductions as prunings of the tree representing the program language.}\vspace{-10pt}
\label{fig:semc}
\end{center}
\end{figure}

Consider the tree depicted in Figure \ref{fig:semc}(i) which corresponds to the language of all traces of a simple program $a \parallel bcd$. Assume that we have $I = \{(b, a), (d, a)\}$. Imagine a (depth-first) traversal of the tree in prefix order starting from the root. Once the left-most branch of the tree is explored, which corresponds to the program run $abcd$, the algorithm explores the right branch at the root. Here,  the algorithm chooses not to explore the run $bacd$, since $(b, a) \in I$ (and therefore $bacd \semc_I abcd$) and $abcd$ has already been explored. This branch is greyed out in Figure \ref{fig:semc}(ii) to indicate that it is pruned. The algorithm continues its exploration and decides to prune $bcda$ since $bcda \semc_I bcad$ and $bcad$ is explored beforehand.

Now let us slightly tweak the algorithm's traversal strategy. At the root, we choose to go right first (instead of left as before). At every other internal node, we do prefix traversal as before.  Now, the algorithm sees $bacd$ first, $bcad$ second, and as before, prunes $bcda$ since $bcda \semc_I bcad$. This is illustrated in Figure \ref{fig:semc}(iii). Then finally, it gets to the leftmost branch from the root and explores $abcd$. Note that $abcd$ cannot be pruned. We have $bacd \semc_I abcd$, but the inverse is not true, that is $abcd \not \semc_I bacd$. The change of the traversal strategy changes the reduction that is acquired.

A particular reduction is parametric on the non-deterministic choices made about which branch to explore first. They determine what program traces are {\em pruned} in favour of others visited before them which are larger with respect to $\semc_I$. Different non-deterministic choices lead to different reductions. Two such reductions are depicted in Figure \ref{fig:semc}(ii,iii) for two different choices of exploration strategy at the root. Note that one can change the exploration strategy at every internal node (with more than one successor) to enumerate more reductions of this particular language. Reductions are then characterized by an assignment $O : \stmt^* \to \linear{\stmt}$ of nodes to linear orderings on $\stmt$, where $(a, b) \in O(\sigma)$ means that at node $\sigma$ (i.e. the node labeled by string $\sigma$ from the root), we explore the child $\sigma a$ after the child $\sigma b$. Each $O$ combined with the semi-independence relation $I$ defines a reduction $\red{P}{I}{O}$ of the program $P$:
\[
  \red{P}{I}{O} = P \setminus \{ \rho a \sigma b \tau \mid \rho, \sigma, \tau \in \stmt^* \land (a, b) \in O(\rho) \land \forall c \in a \sigma  \ldotp (c, b) \in I \}
\]
where smaller (with respect to $\sqsubseteq_I$) strings are pruned away in favour of the larger ones. If program $P$ is upwards closed, then the aggressive pruning defined above is sound:
\begin{lemmarep}\label{lem:semi-pruned}
  For all $O : \stmt^* \to \linear{\stmt}$, if $P$ is upwards-closed then $\red{P}{I}{O} \preceq_I P$.
\end{lemmarep}
\begin{proof}
  First, we define the following order on traces:
  \[
    \sigma a \tau_1 <_O \sigma b \tau_2 \iff (b, a) \in O(x) \land |\tau_1| = |\tau_2|
  \]
  This relation is a variation of the standard lexicographical well-ordering on strings of the same length, and is a well-order as well.

  Assume $\sigma \in P$. It suffices to show that $\sigma \in \floor{\red{P}{I}{O}}_I$. We proceed by induction on $\sigma$ using $<_O$ with the induction hypothesis $\sigma' <_O \sigma \implies \sigma' \in \floor{\red{P}{I}{O}}_I$ for all $\sigma' \in P$.

  If $\sigma \in \red{P}{I}{O}$ then there is nothing left to prove.

  If $\sigma \notin \red{P}{I}{O}$, then by the definition of $\red{P}{I}{O}$ we have $\sigma = \sigma_1 a \sigma_2 b \sigma_3$ for some $\sigma_1, \sigma_2, \sigma_3 \in \stmt^*$ and $a, b \in \stmt$ such that $(a, b) \in O(\sigma_1)$ and $(c, b) \in I$ for all $c \in a\sigma_2$.  Define $\sigma' = \sigma_1 ba \sigma_2 \sigma_3$. Then we have $\sigma = \sigma_1 a \sigma_2 b \sigma_3 \sqsubseteq_I \sigma_1 ba \sigma_2 \sigma_3 = \sigma'$ which implies $\sigma' \in P$ (since $P$ is upwards-closed). Since $(a, b) \in O(\sigma_1)$ we also have $\sigma' <_O \sigma$, and therefore by the inductive hypothesis and transitivity of $\sqsubseteq_I$ we have $\sigma \in \floor{\red{P}{I}{O}}_I$.
\end{proof}

The set of all such reductions for a program and a fixed semi-independence relation $I$ can then be defined by enumerating all such order relations.

\begin{definition}[S-Reduction]\label{def:asemc}
For a sound semi-independence relation $I$ and an upwards closed program $P$, the set of S-reductions of $P$ is defined as
\[\reduce_I(P) = \{ \red{P}{I}{O} \mid O : \stmt^* \to \linear{\stmt} \}.\]
\end{definition}
\begin{wrapfigure}{r}{0.31\textwidth}
\vspace{-12pt}
\begin{center}
\includegraphics[scale=0.19]{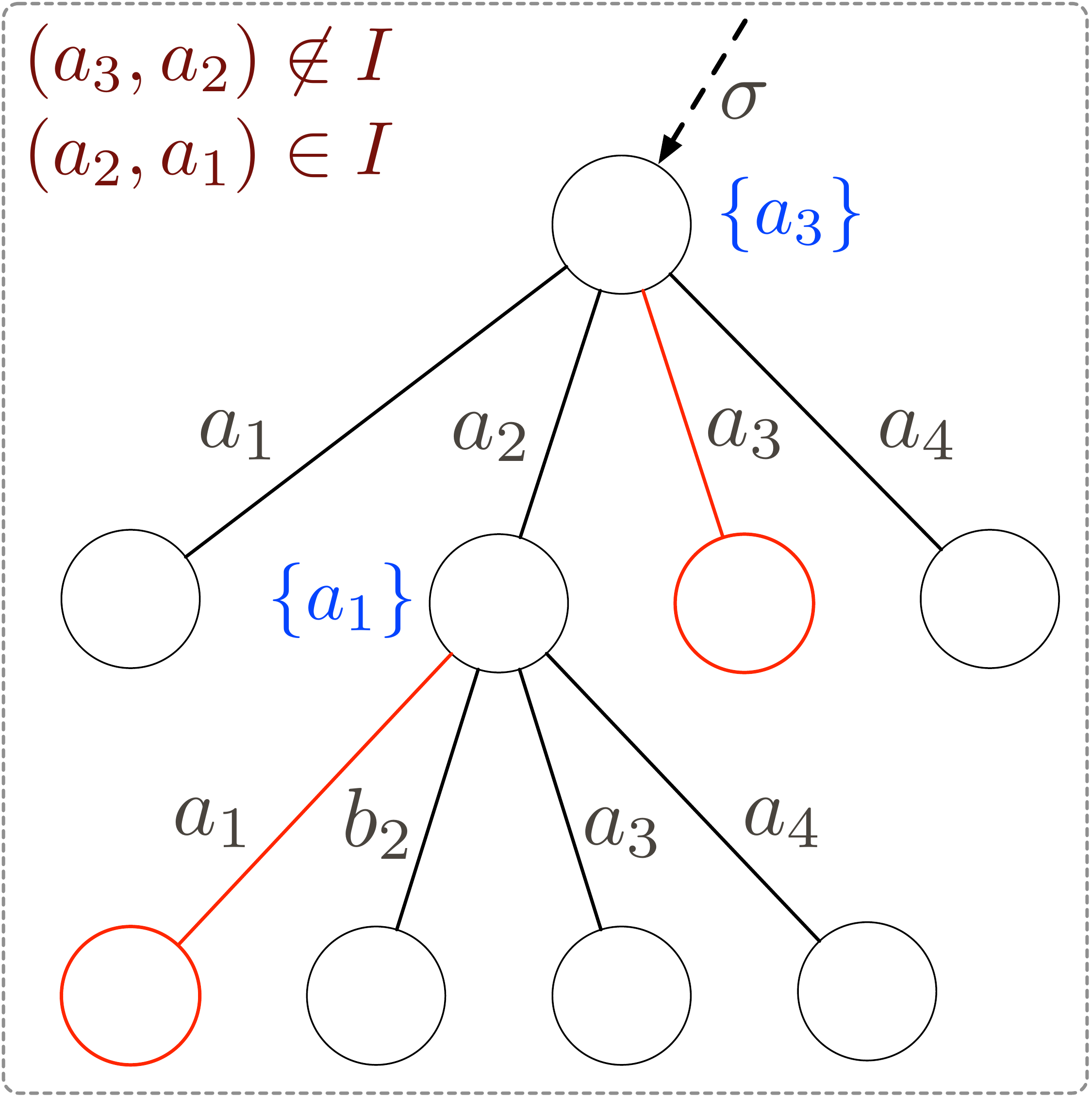}
\end{center}
\vspace{-20pt}
\end{wrapfigure}
The good news is that S-reductions can be effectively represented as the language of an LTA (Looping Tree Automaton) as defined in Section \ref{sec:blta}. The intuition behind the construction of the LTA recognizing $\reduce_I(P)$ is as follows. The state of the LTA keeps track of the set of transitions that can be ignored during the exploration, referred to as {\em \sset sets}. The idea is that the \sset set at the root of the tree is always empty, since nothing can be ignored there. The child node inherits the \sset set of the
parent node, adds to it the transitions that have already been explored from the parent node (which is retrievable from $O(\sigma)$)
and removes from it anything that is not semi-independent on the transition taken from the
parent to the child. Ignored transitions define ignored nodes in a tree in a straightforward manner: a node is not ignored if there is a path of (all) unignored transitions to it from the root.
For example, in the figure on the right, if at node $\sigma$, the transition $a_3$ can be ignored, then it means all the descendents of $\sigma a_3$ are also ignored. If $a_i$'s are traversed in ascending order of $i$'s, then by the time we get to $\sigma a_2$, we have already explored $\sigma a_1$ and its descendants. At $\sigma a_2$, we can ignore $a_1$ in addition to $a_3$ which is already in the \sset set of $\sigma$. However, it is assumed that $(a_2,a_3) \not \in I$. Therefore, we have to remove $a_3$ from the inherited \sset set. Therefore, at $\sigma a_2$, $a_1$ is the only thing that can be ignored.
The LTA effectively accepts all such trees for all possible choices of $O(\sigma)$ at each node $\sigma$ by maintaining these (finite) \sset sets in its state. The full construction, which is inspired by the one given in \cite{cav19} for a symmetric $I$, appears in the proof of the Theorem below:

\begin{theoremrep}\label{thm:reduce-lta}
  For any regular language $P$ and semi-independence relation $I$, the set of S-reductions of $P$ defined by $I$ is recognized by an LTA.
\end{theoremrep}
\begin{proof}
  First, observe that the set $\{ \rho a \sigma b \tau \mid \rho, \sigma, \tau \in \stmt^* \land (a, b) \in O(\rho) \land \forall c \in a \sigma  \ldotp (c, b) \in I \}$ that appears in the definition of $\red{P}{I}{O}$ is equivalent to the set $\ignore_{I, O}$, defined as the smallest set satisfying
  \begin{align*}
    \sigma \in \ignore_{I, O} &\implies \sigma a \in \ignore_{I, O} \\
    a \in \sleep_{I, O}(\sigma) &\implies \sigma a \in \ignore_{I, O}
  \end{align*}
  where $\sleep_{I, O}(\sigma)$ is defined recursively as
  \begin{align*}
    \sleep_{I, O}(\epsilon) &= \emptyset \\
    \sleep_{I, O}(\sigma a) &= (\sleep_{I, O}(\sigma) \cup O(\sigma)(a)) \cap I(a).
  \end{align*}
  The recursive nature of these definitions lend themselves to a simple LTA construction for $\reduce_I(P)$. Let $A_P = (Q, \stmt, \delta, q_0, F)$ be a DFA recognizing $P$. We define our LTA construction $M_P = (Q_P, \stmt, \Delta_P, q_{0P})$ where
  \begin{itemize}
    \item $Q_P = Q \times \bool \times \pow{\stmt}$,
    \item $\begin{aligned}[t]
      \Delta_P &= \{ ((q, \iota, S), B, f) \mid \exists O \in \linear{\stmt} \ldotp \\
      & (B \iff q \in F \land \neg \iota)\ \land \\
      & \forall a \ldotp f(a) = (\delta(q, a), \iota \lor a \in S, (S \cup O(a)) \cap I(a)) \}
    \end{aligned}$,
    \item $q_{0P} = (q_0, \bot, \emptyset)$.
  \end{itemize}
  It follows by a simple inductive proof that any $\delta^*_P : \stmt^* \to Q_P$ is a valid run of $M_P$ iff there exists some $O : \stmt^* \to \linear{\stmt}$ such that $\delta^*_P(\sigma) = (\delta^*(q_0, \sigma), \sigma \in \ignore_{I, O}, \sleep_{I, O}(\sigma))$ for all $\sigma \in \stmt^*$. This implies that $\lang{M_P} = \reduce_I(P)$.
\end{proof}

Since the set of S-reductions is parametric on $I$, it is interesting to explore the connection between two different reduction sets $\reduce_I(P)$ and $\reduce_J(P)$ when $I \subseteq J$. It is tempting to think that $I \subseteq J \implies \forall P \ldotp \reduce_J(P) \subseteq \reduce_I(P)$. The more liberal semi-independence relation, in this case $J$, permits more aggressive prunings and hence produces smaller reductions. But its reductions are not reductions of $I$, specially if $I \subset J$. The following statement is true, which has the same desired positive effect for proof checking:

\begin{propositionrep}\label{prop:red-subsumption}
  Given a program $P$, two semi-independence relations $I$ and $J$, and an ordering function $O : \stmt^* \to \linear{\stmt}$, if $I \subseteq J$ then $\red{P}{J}{O} \subseteq \red{P}{I}{O}$.
\end{propositionrep}
\begin{proof}
  Recall the definitions of $\sleep_{I, O}$ and $\ignore_{I, O}$ from Lemma \ref{thm:reduce-lta}, and that $\red{P}{I}{O} = P \setminus \ignore_{I, O}$. A simple inductive proof gives us $\forall \sigma \ldotp \sleep_{I, O}(\sigma) \subseteq \sleep_{J, O}(\sigma)$. This in turn implies (again by a simple inductive proof) $\forall \sigma \ldotp \ignore_{I, O} \subseteq \ignore_{J, O}$, which implies $\red{P}{J}{O} \subseteq \red{P}{I}{O}$.
\end{proof}
This means that if the program has a proof up to a reduction with a weaker semi-independence relation, it will always have a proof for a reduction according to a stronger semi-independence relation. It also implies in a straightforward manner that these reductions subsume the reductions proposed in \cite{cav19} based on symmetric independence relations.

In the special case where $I$ is symmetric (as is the case in \cite{cav19}), each $L \in \reduce_I(P)$ is guaranteed to be optimal in the sense that the elements of $L$ are pairwise incomparable \cite{cav19}. Unfortunately, this does not hold for a general (non-symmetric) $I$. For example, the language defined by the tree in Figure \ref{fig:semc}(iii) is a strict superset of the language defined by the tree in Figure \ref{fig:semc}(ii). Specifically, the language of the tree in Figure \ref{fig:semc}(iii) contains both $abcd$ and $bacd$, and the latter trace is redundant because $bacd \semc_I abcd$. Therefore, some program reductions defined by $\reduce_I(P)$ contain redundant traces and are non-optimal.

\subsection{Computing a Sound Semi-Independence Relation}\label{sec:cssir}
LTA-representability of the class of S-reductions (Theorem \ref{thm:reduce-lta}) is the key result of Section \ref{sec:semi}. Soundness of S-reductions relies on Lemmas \ref{lem:semi-reduction} and \ref{lem:semi-pruned}.
Lemma \ref{lem:semi-pruned} requires that the program $P$ is upwards-closed with respect to the independence relation $I$ and Lemma \ref{lem:semi-reduction} requires that $I$ is sound. Here, we outline how a relation $I$ that satisfies both criteria can be constructed, with the help of a theorem prover.
Proposition \ref{prop:red-subsumption} implies that one should try to obtain as large of an independence relation as possible in order to maximize the likelihood that there exists a reduction with a proof. One can show that every program $P$ admits a \emph{maximal semi-independence relation} $I_P$.

The (Brzozowski) derivative of a language $P$ and a string $\sigma$ is defined as
\[\sigma^{-1}P = \{\tau \in \stmt^* \mid \sigma \tau \in P\}. \]

\begin{theoremrep}\label{thm:maximal-largest}
The relation $I_P$ defined as
$I_P = \{ (a, b) \mid a \ne b \land \forall \sigma \ldotp (\sigma ab)^{-1}P \subseteq (\sigma ba)^{-1}P \}$
is the largest (with respect to $\subseteq$) semi-independence relation such that  $P = \ceil{P}_{I_P}$ (upwards closedness of $P$).
\end{theoremrep}
\begin{proof}
  First we show $P = \ceil{P}_{I_P}$. Clearly $P \subseteq \ceil{P}_{I_P}$, so it suffices to show $\ceil{P}_{I_P} \subseteq P$.

  Assume $\sigma \in \ceil{P}_{I_P}$. Since $\sigma$ is in the upwards closure of $P$, there must exist something in $P$ that is below $\sigma$, so we obtain some $\tau \in P$ such that $\tau \sqsubseteq_I \sigma$. Then $\tau$ can be obtained from $\sigma$ by performing some finite number of swaps $n$; we shall use $\tau \sqsubseteq_I^n \sigma$ to denote this. We proceed by induction on $n$, with the inductive hypothesis $\tau' \sqsubseteq_I^{n - 1} \sigma \implies \sigma \in \ceil{P}_{I_P}$ for all $\tau' \in P$ whenever $n > 0$.

  If $n = 0$, then $\tau = \sigma$, so clearly $\sigma \in P$.

  If $n > 0$, then $\tau = \tau_1 ab \tau_2$ for some $\tau_1, \tau_2 \in \stmt^*$ and $a, b \in \stmt$ such that $(a, b) \in I_P$ and $\tau_1 ba \tau_2 \sqsubseteq_{I_P}^{n - 1} \sigma$. By the definition of $I_P$ we have $\tau_1 ba \tau_2 \in P$ from $\tau_1 ab \tau_2 \in P$, and by the inductive hypothesis we have $\sigma \in P$.

  Next, we show that $I_P$ is maximal. It suffices to show that $P$ is not upwards closed for any semi-independence relation $I$ that includes a pair $(a, b) \in I$ such that $(\sigma ab)^{-1} P \nsubseteq (\sigma ba)^{-1} P$ for some $\sigma \in P$. By the definition of the derivative there exists some $\tau$ such that $\sigma ab \tau \in P$ and $\sigma ba \tau \notin P$. Then $\sigma ab \tau \sqsubseteq_I \sigma ba \tau$, which violates upwards closedness.
\end{proof}
When $P$ is regular (as is the case for all of our input programs), it is possible to construct $I_P$ directly, as the proposition $\forall \sigma \ldotp (\sigma ab)^{-1}P \subseteq (\sigma ba)^{-1}P$ is equivalent to a subsumption relation on states of the  DFA recognizing $P$ \cite{diekert1995book}.
\begin{theoremrep}
  If $P$ is regular, then $I_P$ is computable.
\end{theoremrep}
\begin{proof}
  Let $A = (Q, \stmt, \delta, q_0, F)$ be the minimal DFA representing $P$. We define $A_q$ to be the DFA obtained by replacing $q_0$ with $q$ in $A$. Then we have
  \begin{align*}
    (\forall \sigma \ldotp (\sigma ab)^{-1} P \subseteq (\sigma ba)^{-1} P)
    &\iff (\forall \sigma, \tau \ldotp \sigma ab \tau \in P \implies \sigma ba \tau \in P) \\
    &\iff (\forall \sigma, \tau \ldotp \delta^*(q_0, \sigma ab \tau) \in F \implies \delta^*(q_0, \sigma ba \tau) \in F) \\
    &\iff (\forall \sigma, \tau \ldotp \delta^*(\delta^*(q_0, \sigma ab), \tau) \in F \implies \delta^*(\delta^*(q_0, \sigma ba), \tau) \in F) \\
    &\iff (\forall q, \tau \ldotp \delta^*(\delta^*(q, ab), \tau) \in F \implies \delta^*(\delta^*(q, ba), \tau) \in F) \\
    &\iff (\forall q \ldotp \lang{A_{\delta^*(q, ab)}} \subseteq \lang{A_{\delta^*(q, ba)}})
  \end{align*}
  Since regular language inclusion is decidable, it follows that we can compute $I_P$ by iterating over all possible pairs of statements.
\end{proof}
$I_P$ may be unsound. One can always obtain a \emph{maximal sound semi-independence relation} by removing from $I_P$ all statements $a$ and $b$ that do not satisfy $\sem{ab} \subseteq \sem{ba}$. This last step needs to be performed by making calls to a theorem prover. Note that since $a$ and $b$ are program statements, the computation of this relation takes place once at the beginning of the verification process by making a quadratic number (in the program size) of calls to a solver.

\section{Contextual Reductions}\label{sec:context}
\begin{toappendix}
    \label{app:context}
\end{toappendix}
We introduce the notion of {\em context} for program reductions through the definition of a {\em contextual semi-independence} relation. The independence relation is strengthened through the consideration of the context information, where statements can be declared (semi-) commutative only in some contexts. Context is typically considered to be a state (or a set of states) from which the transitions are being considered \cite{KatzP92,GodefroidP93}. We propose a different notion of context which is more useful in our language-theoretic setting, where the {\em history} of the trace is used as context.

Concretely, a \emph{contextual semi-independence relation} is a function $\mathcal{I} : \stmt^* \to \pow{\stmt \times \stmt}$ from traces to irreflexive relations. Intuitively $(a,b) \in \mathcal{I}(\sigma)$ should hold for statements $a$ and $b$ and a program trace $\sigma$ where $a$ can be swapped with $b$ in context $\sigma$, that is $\sem{\sigma ab} \subseteq \sem{\sigma ba}$. Note that contextual semi-independence subsumes normal semi-independence, which can be considered as the special case of a constant function; i.e. the same independence relation is assigned to all contexts.

Define $\semc_\mathcal{I}$ to be the smallest preorder satisfying $\sigma ab\rho \semc_\mathcal{I} \sigma ba\rho$ for all $\sigma, \rho \in \stmt^*$ and $(a, b) \in \mathcal{I}(\sigma)$. Upwards and downwards closure and closedness are defined as before. We say $\mathcal{I}$ is \emph{sound} if $\sem{\sigma ab} \subseteq \sem{\sigma ba}$ for all $\sigma  \in \stmt^*$ and $(a, b) \in \mathcal{I}(\sigma )$.

\begin{example}
Recall the example of Figure \ref{fig:me2}, where we discussed the idea of contextual commutativity of {\tt inc()} and {\tt dec()} at the high level in Section \ref{sec:example}. Concretely, $(\mbox{{\tt inc()}}, \mbox{{\tt dec()}}) \in \mathcal{I}(\sigma)$ and $(\mbox{{\tt dec()}},\mbox{{\tt inc()}}) \in \mathcal{I}(\sigma)$, for all $\sigma$ where the number of {\tt inc()} statements is strictly larger than the number of {\tt dec()} statements in $\sigma$.
\end{example}

Since our contexts are defined language-theoretically, the definition of $\downarrow$ can be naturally extended to support contexts. Define
\[
   \red{P}{\mathcal{I}}{O} = P \setminus \{ \sigma a\rho b\upsilon \mid \sigma, \rho, \upsilon \in \stmt^* \land (a, b) \in O(\sigma) \land \forall \tau c \preceqdot a\rho \ldotp (c, b) \in \mathcal{I}(\sigma \tau) \}
\]
where $\preceqdot$ is the prefix relation on strings. Similar to the definition of semi-commutative reductions, strings are soundly pruned from the program language to obtain each reduction $\red{P}{\mathcal{I}}{O}$, where $O$ is an order that determines the exploration strategy for the particular reduction. The set of all reductions is then defined as
\[
    \creduce_\mathcal{I}(P) = \{ \red{P}{\mathcal{I}}{O} \mid O : \stmt^* \to \linear{\stmt} \}.
\]
When $\mathcal{I}$ is a constant function, which makes the contextual relation collapse into the standard semi-independence relation of Definition \ref{def:semc}, $\creduce_\mathcal{I}(P)$ is representable as an LTA (by Theorem \ref{thm:reduce-lta}). This does not hold true for a general $\mathcal{I}$. Since the goal of this paper is the development of algorithms for enumerating reductions effectively, we are strictly interested in cases where for a given $\mathcal{I}$, the set of reductions $\creduce_\mathcal{I}(P)$ is LTA representable.

\subsection{A Representable Class of Contextual Reductions}\label{sec:repcom}
A contextual semi-independence relation $\mathcal{I} : \stmt^* \to \pow{\stmt \times \stmt}$ can be alternatively viewed as an infinite tree labelled by a standard semi-independence relation. Thus we call $\mathcal{I}$ \emph{regular} if it corresponds to a regular tree. An infinite tree is {\em regular} iff it contains a finite number of unique subtrees. Equivalently, an infinite tree is regular iff it can be generated by a modified DFA with states marked with arbitrary labels (in our case, semi-independence relations) instead just being labelled as final or non-final. Since  $O: \stmt^* \to \linear{\stmt}$ can be viewed as an infinite tree, its regularity can be accordingly defined. Program reductions induced by regular contextual independence relations are LTA-representable:

\begin{theoremrep}\label{thm:reduce-lta-2}
    If $\mathcal{I}$ is regular, then $\creduce_\mathcal{I}(P)$ is representable as an LTA.
\end{theoremrep}
\begin{proof}
  First, observe that the set $\{ \sigma a\rho b\upsilon \mid \sigma, \rho, \upsilon \in \stmt^* \land (a, b) \in O(\sigma) \land \forall \tau c \preceqdot a\rho \ldotp (c, b) \in \mathcal{\mathcal{I}}(\sigma \tau) \}$ that appears in the definition of $\red{P}{\mathcal{I}}{O}$ is equivalent to the set $\ignore_{\mathcal{I}, O}$, defined as the smallest set satisfying
  \begin{align*}
    \sigma \in \ignore_{\mathcal{I}, O} &\implies \sigma a \in \ignore_{\mathcal{I}, O} \\
    a \in \sleep_{\mathcal{I}, O}(\sigma) &\implies \sigma a \in \ignore_{\mathcal{I}, O}
  \end{align*}
  where $\sleep_{\mathcal{I}, O}(\sigma)$ is defined recursively as
  \begin{align*}
    \sleep_{\mathcal{I}, O}(\epsilon) &= \emptyset \\
    \sleep_{\mathcal{I}, O}(\sigma a) &= (\sleep_{\mathcal{I}, O}(\sigma) \cup O(\sigma)(a)) \cap \mathcal{I}(\sigma)(a).
  \end{align*}
  The recursive nature of these definitions lend themselves to a simple LTA construction for $\creduce_\mathcal{I}(P)$. Let $A_P = (Q_P, \stmt, \delta_P, q_{0P}, F_P)$ and $A_\mathcal{I} = (Q_{\mathcal{I}}, \stmt, \delta_{\mathcal{I}}, q_{0{\mathcal{I}}}, F_{\mathcal{I}})$ be automata recognizing $P$ and $\mathcal{I}$, respectively. We define our LTA construction $M_{P\mathcal{I}} = (Q_{P\mathcal{I}}, \stmt, \Delta_{P\mathcal{I}}, q_{0P\mathcal{I}})$ where
  \begin{itemize}
    \item $Q_{P\mathcal{I}} = Q_P \times Q_\mathcal{I} \times \bool \times \pow{\stmt}$,
    \item $\begin{aligned}[t]
      \Delta_P &= \{ ((q_P, q_\mathcal{I}, \iota, S), B, f) \mid \exists O \in \linear{\stmt} \ldotp \\
      & (B \iff q_P \in F_P \land \neg \iota)\ \land \\
      & \forall a \ldotp f(a) = (\delta_P(q_P, a), \delta_\mathcal{I}(q_\mathcal{I}, a), \iota \lor a \in S, (S \cup O(a)) \cap F_\mathcal{I}(q_\mathcal{I})(a)) \}
    \end{aligned}$,
    \item $q_{0P\mathcal{I}} = (q_{0P}, q_{0\mathcal{I}}, \bot, \emptyset)$.
  \end{itemize}
  It follows by a simple inductive proof that any $\delta^*_{P\mathcal{I}} : \stmt^* \to Q_{P\mathcal{I}}$ is a valid run of $M_{P\mathcal{I}}$ iff there exists some $O : \stmt^* \to \linear{\stmt}$ such that $\delta^*_{P\mathcal{I}}(\sigma) = (\delta_P^*(q_{0P}, \sigma), \delta_\mathcal{I}^*(q_{0\mathcal{I}}, \sigma), \sigma \in \ignore_{\mathcal{I}, O}, \sleep_{\mathcal{I}, O}(\sigma))$ for all $\sigma \in \stmt^*$. This implies that $\lang{M_{P\mathcal{I}}} = \creduce_\mathcal{I}(P)$.
\end{proof}
Similar to the semi-commutative case (stated in Theorem \ref{thm:maximal-largest}), every program $P$ has a \emph{maximal contextual semi-independence relation} $\mathcal{I}_P$, defined as
\[
  \mathcal{I}_P(\sigma) = \{ (a, b) \mid a \ne b \land (\sigma ab)^{-1}P \subseteq (\sigma ba)^{-1}P \}.
\]
One can naturally lift $\subseteq$ to functions, where $\mathcal{I}_1 \subseteq \mathcal{I}_2$ iff $\forall \sigma \ldotp \mathcal{I}_1(\sigma) \subseteq \mathcal{I}_2(\sigma)$. This provides an order on the set of contextual relations with respect to which one can define maximality. This leads us to the contextual analog of Theorem \ref{thm:maximal-largest}:
\begin{theoremrep}\label{thm:maximal-largest-context}
    $\mathcal{I}_P$ is the largest (with respect to $\subseteq$) semi-independence relation satisfying $P = \ceil{P}_{\mathcal{I}_P}$.
\end{theoremrep}
\begin{proof}
  First we show $P = \ceil{P}_{\mathcal{I}_P}$. Clearly $P \subseteq \ceil{P}_{\mathcal{I}_P}$, so it suffices to show $\ceil{P}_{\mathcal{I}_P} \subseteq P$.

  Assume $\sigma \in \ceil{P}_{\mathcal{I}_P}$. Since $\sigma$ is in the upwards closure of $P$, there must exist something in $P$ that is below $\sigma$, so we obtain some $\tau \in P$ such that $\tau \sqsubseteq_\mathcal{I} \sigma$. Then $\tau$ can be obtained from $\sigma$ by performing some finite number of swaps $n$; we shall use $\tau \sqsubseteq_\mathcal{I}^n \sigma$ to denote this. We proceed by induction on $n$, with the inductive hypothesis $\tau' \sqsubseteq_\mathcal{I}^{n - 1} \sigma \implies \sigma \in \ceil{P}_{\mathcal{I}_P}$ for all $\tau' \in P$ whenever $n > 0$.

  If $n = 0$, then $\tau = \sigma$, so clearly $\sigma \in P$.

  If $n > 0$, then $\tau = \tau_1 ab \tau_2$ for some $\tau_1, \tau_2 \in \stmt^*$ and $a, b \in \stmt$ such that $(a, b) \in \mathcal{I}_P(\tau_1)$ and $\tau_1 ba \tau_2 \sqsubseteq_{\mathcal{I}_P}^{n - 1} \sigma$. By the definition of $\mathcal{I}_P$ we have $\tau_1 ba \tau_2 \in P$ from $\tau_1 ab \tau_2 \in P$, and by the inductive hypothesis we have $\sigma \in P$.

  Next, we show that $\mathcal{I}_P$ is maximal. It suffices to show that $P$ is not upwards closed for any contextual semi-independence relation $I$ that includes a pair $(a, b) \in \mathcal{I}(\sigma)$ such that $(\sigma ab)^{-1} P \nsubseteq (\sigma ba)^{-1} P$ for some $\sigma \in P$. By the definition of the derivative there exists some $\tau$ such that $\sigma ab \tau \in P$ and $\sigma ba \tau \notin P$. Then $\sigma ab \tau \sqsubseteq_\mathcal{I} \sigma ba \tau$, which violates upwards closedness.
\end{proof}
When $P$ is a regular language, $\mathcal{I}_P$ is a computable function. In fact, a stronger result holds:
\begin{theoremrep}
    If $P$ is regular then so is $\mathcal{I}_P$.
\end{theoremrep}
\begin{proof}
    Let $A = (Q, \stmt, \delta, q_0, F)$ be a DFA for $P$. Define $A' = (Q, \stmt, \delta, q_0, F')$ where $F' : \stmt^* \to \pow{\stmt \times \stmt}$ is defined as
    \[
        F'(q) = \{ (a, b) \mid a \ne b \land \lang{A_{\delta^*(q, ab)}} \subseteq \lang{A_{\delta^*(q, ba)}} \}.
    \]
    Then
    \begin{align*}
        (a, b) \in \mathcal{I}_P(\sigma)
        &\iff a \ne b \land (\sigma ab)^{-1}P \subseteq (\sigma ba)^{-1}P \\
        &\iff a \ne b \land (\forall \tau \ldotp \sigma ab \tau \in P \implies \sigma ba \tau \in P) \\
        &\iff a \ne b \land (\forall \tau \ldotp \delta^*(q_0, \sigma ab \tau) \in F \implies \delta^*(q_0, \sigma ba \tau) \in F) \\
        &\iff a \ne b \land (\forall \tau \ldotp \delta^*(q_0, \sigma ab \tau) \in F \implies \delta^*(q_0, \sigma ba \tau) \in F) \\
        &\iff a \ne b \land (\forall \tau \ldotp \delta^*(\delta^*(q_0, \sigma ab), \tau) \in F \implies \delta^*(\delta^*(q_0, \sigma ba), \tau) \in F) \\
        &\iff a \ne b \land A_{\delta^*(\delta^*(q_0, \sigma), ab)} \subseteq A_{\delta^*(\delta^*(q_0, \sigma), ba)} \\
        &\iff (a, b) \in F'(\delta^*(q_0, \sigma))
    \end{align*}
    which is the acceptance condition for $A'$.
\end{proof}
As in the semi-commutative case, there is no guarantee that $\mathcal{I}_P$ is sound, and one may obtain a maximal sound contextual semi-independence relation $\mathcal{I}^\mathrm{sound}_P$ by removing the unsound elements:
\[
    \mathcal{I}^\mathrm{sound}_P(\sigma) = \mathcal{I}_P(\sigma) \setminus \{ (a, b) \mid \sem{\sigma ab} \nsubseteq \sem{\sigma ba} \}.
\]
Unlike the semi-commutative case, where any semi-commutative relation defines an LTA-representable set of reductions, there is no guarantee that $\creduce_{\mathcal{I}^\mathrm{sound}_P}(P)$ is LTA-representable.
\begin{theoremrep}
    The set $\creduce_{\mathcal{I}^\mathrm{sound}_P}(P)$ generally cannot be represented by an LTA.
\end{theoremrep}
\begin{proof}
    We reduce representability of $\creduce_{\mathcal{I}^\mathrm{sound}_P}(P)$ to safety. For a contradiction, assume $\creduce_{\mathcal{I}^\mathrm{sound}_P}(P)$ is representable by an LTA for any regular $P$.

    Let $P'$ be any regular program, and let $a$ and $b$ be two statements that do not appear in $P'$, are always enabled, and only soundly commute in an inconsistent context, i.e. $\sem{a}$ and $\sem{b}$ are total and $a$ and $b$ only commute in context $\sigma$ if $\sigma$ is infeasible. For example, one can take statements $x \gets 0$ and $x \gets 1$.

    Let $P = P' \cdot \{ ab, ba \}$. By our initial assumption, $\creduce_{\mathcal{I}^\mathrm{sound}_P}(P)$ is representable by an LTA. By definition, $\creduce_{\mathcal{I}^\mathrm{sound}_P}(P)$ is non-empty. Every non-empty LTA contains a regular language \cite{cav19}, so we obtain a regular language $R \in \creduce_{\mathcal{I}^\mathrm{sound}_P}(P)$. Since $R$ is a reduction of $P$ and a subset of $P$, it follows that $P$ is sound iff $R$ is sound.

    Every trace in $R$ ends in either $ab$ or $ba$. If $P$ is sound, then for any $\sigma \in P$ we cannot have both $\sigma ab \in R$ and $\sigma ba \in R$: $\sigma$ is infeasible, which means $a$ and $b$ are independent in context $\sigma$, which means that at least one of these strings will be pruned (depending on whether $a$ or $b$ is explored first at $\sigma$).

    Conversely, if $P$ is unsound then so is $R$, and there must exist a feasible trace $\sigma ab \in R$. Since $a$ and $b$ are always enabled, it follows that $\sigma$ is also feasible. Since $\sigma$ is feasible, it follows that $a$ and $b$ do not commute in context $\sigma$, so $R$ must also include $\sigma ba$.

    Thus $P$ is unsafe iff $R$ contains a pair of traces $\sigma ab$ and $\sigma ba$. Since $R$ is regular, this can easily be checked by examining the states of the DFA accepting $R$. Thus we have a decision procedure for safety, which is not possible.
\end{proof}
This makes $\mathcal{I}^\mathrm{sound}_P$ unsuitable for use in automated verification. Fortunately, we have a solution for this problem.
Recall that the ultimate goal is to find some reduction $\red{P}{\mathcal{I}}{O}$ that can be proven safe, and to this end, it is not necessary that $\mathcal{I}$ be maximal. It is difficult (if not impossible), without knowing anything about the proof, to make a correct choice about $\mathcal{I}$ in advance. Therefore, we can be clever and handle the choice of $\mathcal{I}$ in the same way that we handle the choice of a particular exploration order $O$ for a reduction: we construct the set of all possible $\red{P}{\mathcal{I}}{O}$ over all possible $\mathcal{I}$ and $O$. It will be left to the verification algorithm to {\em discover} the right choices for both $\mathcal{I}$ and $O$.

\begin{wrapfigure}{r}{0.34\textwidth}
\vspace{-10pt}
\begin{center}
\includegraphics[scale=0.2]{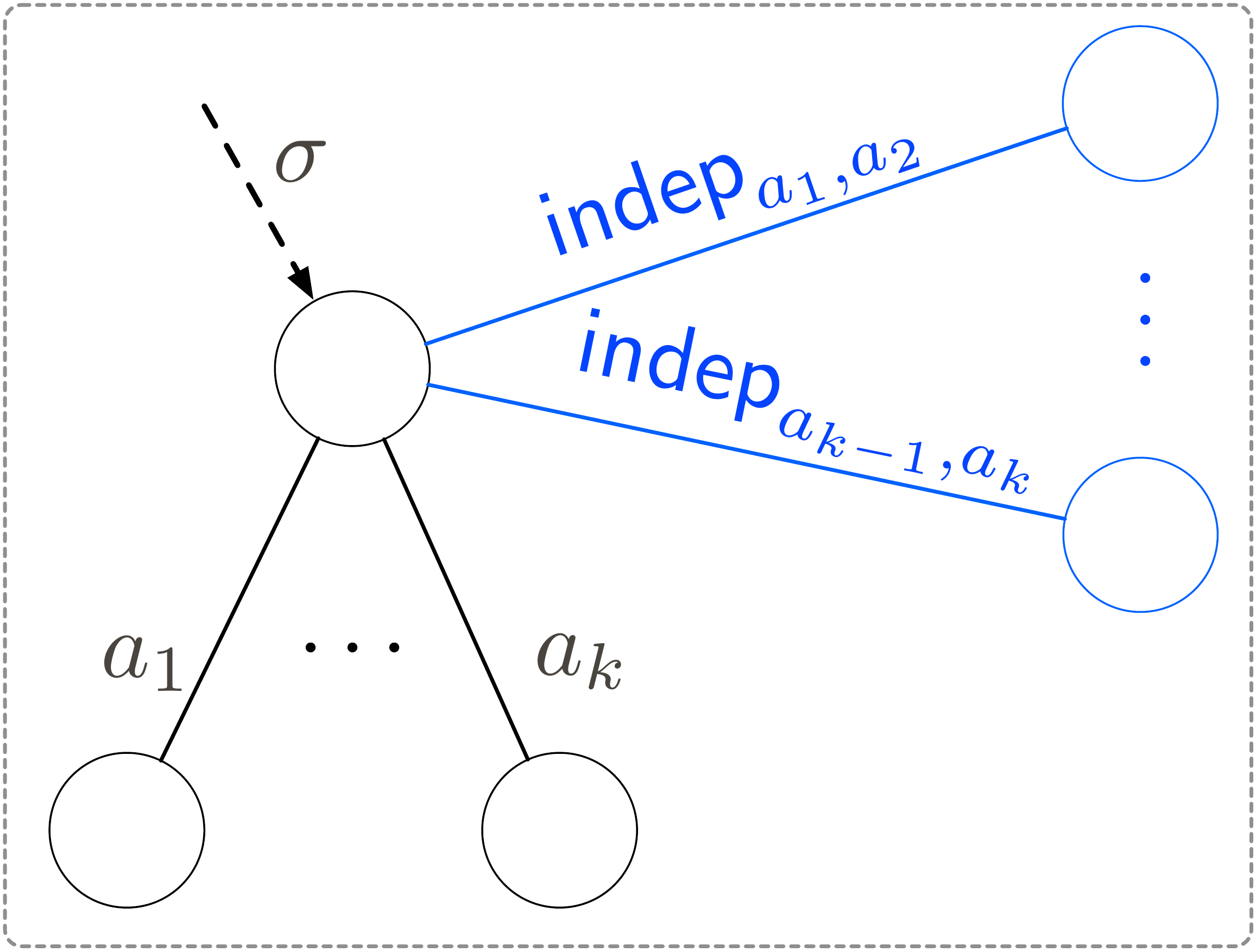}
\end{center}
\vspace{-10pt}
\end{wrapfigure}Since not all independence relations are sound, we ensure our reductions are valid by adding additional \emph{soundness constraints} to each reduction in the form of additional traces that can only be proven correct if the underlying independence relation is sound. This is done via an additional set of \emph{independence statements} $\stmt_\mathrm{indep} = \{  \mathrm{indep}_{a, b} \mid a, b \in \stmt \}$ with the semantics $\sem{\mathrm{indep}_{a, b}} = \sem{ab} \setminus \sem{ba}$. Intuitively, each $\mathrm{indep}_{a, b}$ is infeasible iff it is executed in a state where statement $a$ commutes to the right of $b$. As illustrated on the right, for every node $\sigma$, and each pair of outgoing transitions $a_i$ and $a_j$, a new independence transition with the label $\mathrm{indep}_{a_i, a_j}$ is added to a new fresh state. The states/transitions illustrated in blue are the new additions.
The set of soundness constraints for a particular independence relation $\mathcal{I} : \stmt^* \to \pow{\stmt \times \stmt}$ is then defined as
\[
    \mathrm{sound}(\mathcal{I}) = \{ \sigma  \cdot \mathrm{indep}_{a, b} \mid (a, b) \in \mathcal{I}(\sigma ) \}.
\]
Intuitively, this corresponds to unreachability of all newly added (blue) states in the schematic figure above, which is formalized in the lemma below:
\begin{lemmarep}\label{lem:sound-safe}
    $\mathcal{I}$ is sound iff $\sem{\mathrm{sound}(\mathcal{I})} = \emptyset$.
\end{lemmarep}
\begin{proof}
    Assume $\mathcal{I}$ is sound and let $\sigma \cdot \mathrm{indep}_{a, b}$ be any trace in $\mathrm{sound}(\mathcal{I})$. Then
    \begin{align*}
        \sem{\sigma \cdot \mathrm{indep}_{a, b}}
        &= \sem{\sigma} \circ \sem{\mathrm{indep}_{a, b}} \\
        &= \sem{\sigma} \circ (\sem{ab} \setminus \sem{ba}) \\
        &= (\sem{\sigma} \circ \sem{ab}) \setminus (\sem{\sigma} \circ \sem{ba}) \\
        &= \sem{\sigma ab} \setminus \sem{\sigma ba} \\
        &= \emptyset,
    \end{align*}
    so soundness of $\mathcal{I}$ implies safety of $\mathrm{sound}(\mathrm{I})$.

    For the other direction, assume $\sigma \in \stmt^*$ and $a, b \in \stmt$ such that $\sem{\sigma \cdot \mathrm{indep}_{a, b}} = \emptyset$. By the above derivation we have $\sem{\sigma ab} \setminus \sem{\sigma ba} = \emptyset$, which implies $\sem{\sigma ab} \subseteq \sem{\sigma ba}$.
\end{proof}
At this point, we can claim $P$ is safe if \emph{both} $\red{P}{\mathcal{I}}{O}$ and $\mathrm{sound}(\mathcal{I})$ are safe. Since two programs are safe iff their union is safe, we can treat $\red{P}{\mathcal{I}}{O}$ and $\mathrm{sound}(\mathcal{I})$ as a single reduction by taking their union.

\begin{theoremrep}
    For any independence relation $\mathcal{I} : \stmt^* \to \pow{\stmt \times \stmt}$ and upwards-closed program $P$, $\red{P}{\mathcal{I}}{O} \cup \mathrm{sound}(\mathcal{I}) \preceq P$.
\end{theoremrep}
\begin{proof}
    Follows trivially from Lemma \ref{lem:sound-safe} and soundness of C-reductions.
\end{proof}
Finally, we are ready to define our set of contextual reductions.

\begin{definition}[C-Reductions]\label{def:conc}
  Given a program $P$, the set of C-reductions of the program is defined as:
\[
    \creduce^*(P) = \{ \red{P}{\mathcal{I}}{O}\cup \mathrm{sound}(\mathcal{I}) \mid
        O : \stmt^* \to \linear{\stmt},
        \mathcal{I} \subseteq \mathcal{I}_P \}.
\]
\end{definition}
C-reductions, like S-reductions, are effectively representable for algorithmic verification:

\begin{theoremrep}\label{thm:cred-lta}
    For any regular program $P$ the set of C-reductions (i.e. $\creduce^*(P)$) of $P$ is recognized by an LTA.
\end{theoremrep}
\begin{proof}
    Let $A_P = (Q_P, \stmt, \delta_P, q_{0P}, F_P)$ and $A_{\mathcal{I}_P} = (Q_{\mathcal{I}_P}, \stmt, \delta_{\mathcal{I}_P}, q_{0{\mathcal{I}_P}}, F_{\mathcal{I}_P})$ be automata recognizing $P$ and $\mathcal{I}_P$, respectively. We define our LTA construction $M_{P^*} = (Q_{P^*}, \stmt, \Delta_{P^*}, q_{0P^*})$ where
    \begin{itemize}
      \item $Q_{P^*} = (Q_P \times Q_{\mathcal{I}_P} \times \bool \times \pow{\stmt}) \cup \bool$,
      \item $\begin{aligned}[t]
        \Delta_P =\ & \{ (B, B, \lambda a \ldotp \bot) \} \\
              \cup\ &
        \begin{aligned}[t]
                \{ & ((q_P, q_{\mathcal{I}_P}, \iota, S), B, f) \mid \exists I \subseteq F_{\mathcal{I}_P}(q_{\mathcal{I}_P}), O \in \linear{\stmt} \ldotp \\
        & (B \iff q_P \in F_P \land \neg \iota)\ \land \\
        & (\forall a, b \ldotp f(\mathrm{indep}_{a, b}) = ((a, b) \in I))\ \land \\
        & (\forall a \notin \stmt_\mathrm{indep} \ldotp f(a) = (\delta_P(q_P, a), \delta_{\mathcal{I}_P}(q_{\mathcal{I}_P}, a), \iota \lor a \in S, (S \cup O(a)) \cap I(a))) \}
      \end{aligned}
      \end{aligned}$,
      \item $q_{0P^*} = (q_{0P}, q_{0{\mathcal{I}_P}}, \bot, \emptyset)$.
    \end{itemize}
    It follows by a simple inductive proof that any $\delta^*_{P^*} : \stmt^* \to Q_{P^*}$ is a valid run of $M_{P^*}$ iff there exists some $\mathcal{I} : \stmt^* \to \pow{\stmt \times \stmt}$ and $O : \stmt^* \to \linear{\stmt}$ such that $\delta^*_{P^*}(\sigma) = (\delta_P^*(q_{0P}, \sigma), \delta_{\mathcal{I}_P}^*(q_{0{\mathcal{I}_P}}, \sigma), \sigma \in \ignore_{\mathcal{I}, O}, \sleep_{\mathcal{I}, O}(\sigma))$, $\delta^*_{P^*}(\sigma \cdot \mathrm{indep}_{a, b}) = \top$, and $\delta^*_{P^*}(\sigma \cdot \mathrm{indep}_{a, b} \cdot \tau) = \bot$ for all $\sigma \in \stmt^*$, $\tau \in (\stmt \cup \stmt_\mathrm{indep})^+$, and $a, b \in \stmt$. This implies that $\lang{M_{P^*}} = \creduce^*(P)$.
\end{proof}
Note that since the LTA represents the set of all reductions with all possible choices of $\mathcal{I}$, it includes the reductions with the specific choice of the maximal sound contextual relation $\mathcal{I}^\mathrm{sound}_P$. Therefore, reductions based on $\mathcal{I}^\mathrm{sound}_P$ will be considered for the proof without the need for them to be captured by an LTA as a single set.

\subsection{Finite Programs}
There is research \cite{WangCGY09} that focuses on reductions in the context of bounded model checking (i.e. bug finding) for concurrent programs. It is therefore worthwhile to mention that for this special class of programs, where the program language includes only finitely many traces, an appropriate regular reduction always exists.
\begin{theoremrep}\label{thm:finite}
    For any finite $P$, there exists a sound regular independence relation $\mathcal{I} \subseteq \mathcal{I}_P$ such that $\creduce_\mathcal{I}(P) = \creduce_{\mathcal{I}_P}(P)$.
\end{theoremrep}
\begin{proof}
    Any trace $\tau$ that is not a prefix of any trace in $P$ is irrelevant as far as independence is concerned. An independence relation may choose to declare any and all statements independent at $\tau$ without affecting the set of traces to be pruned. There is nothing to prune anyways. Thus we define $\mathcal{I}$ to be the restriction of $\mathcal{I}_P$ to $P$:
    \[
        \mathcal{I}(\sigma) = \{ (a, b) \in \mathcal{I}_P(\sigma) \mid \exists \tau \ldotp \sigma\tau \in P \}.
    \]
    While $\mathcal{I}(\sigma)$ may not have a finite representation, it is still computable (assuming the set of statements are semantically within a decidable logic). Since $P$ is finite, we may calculate exactly what independencies should hold at each trace in $P$. An automaton accepting $\mathcal{I}(\sigma)$ simply has to check whether $\sigma \in P$.
\end{proof}
While the ``ideal'' independence relation $\mathcal{I}^\mathrm{sound}_P$ defined previously may not be regular (and therefore may not satisfy the precondition of Theorem \ref{thm:reduce-lta-2}), Theorem \ref{thm:finite} implies that we can always construct another regular independence relation that produces equivalent reductions. This implies that completeness with respect to reductions can be achieved for bounded programs, while it is not achievable for general (unbounded) programs. \anthony{I think we need to specify \emph{which} reductions we are talking about, since there are 3-4 different notions flying around at this point.}




\section{Relationship to Known Reduction Techniques}\label{sec:relations}
Now that we have introduced C-reductions, it is only natural to ask if they {\em subsume} some well-understood and widely used reduction techniques specific to certain program classes. To this end, we investigate the relation between C-reductions and the two known approaches of  Lipton \cite{Lipton75} and Existential Boundedness \cite{Genest07}.

\subsection{Lipton's Atomic Block Reductions}
In Lipton's reductions, semi-commutativity properties of statements are used as sufficient conditions to infer {\em atomic blocks}. Declaring a block of code atomic (and having it executed without interruption) has the effect of reducing the number of thread interleavings that must be proved correct.

Lipton's condition is based on \emph{left movers} and \emph{right movers}. A left mover (respectively right mover) is a statement that soundly commutes to the left (respectively right) of every statement in every other thread. A sequence of statements $a_1 \ldots a_n b c_1 \ldots c_m$ may be declared atomic if each $a_i$ is a right mover and each $c_j$ is a left mover. First, note that Lipton's original definition for left/right-movers was a {\em contextual} definition. He defined an action to be a left/right-mover if it left/right commutes from all concrete reachable program contexts. For a program with variables ranging over infinite data domains, this implies infinitely many possible contexts. Therefore, a corresponding contextual commutativity relation will be incomputable in general. Given a general contextual left/right-mover specification, checking the soundness of it according to Lipton's original contextual definition is undecidable since determining the set of reachable states is undecidable in general. This is perhaps why all instances of usage of Lipton's reductions in the literature are non-contextual. One can view our C-reductions as an attempt to provide a decidable approximation of this original definition for program safety verification.

Let us now compare our (non-contextual) S-reductions against the commonly used non-contextual definition of Lipton's reductions. First, observe that Lipton's reductions are not optimal, even if maximal blocks of atomic codes are inferred. For example, consider a {\em disjoint} parallel program $a_1 a_2 \parallel b_1 b_2$.  Since every statement is both a left and a right mover, one can reduce the program to $A \parallel B$ where $A = a_1 a_2$ and $B = b_1 b_2$ are atomic blocks. Note that this reduction includes two execution $AB$ and $BA$ which are equivalent. Therefore, one is redundant. The set of S-reductions of this program will include reductions with no redundancies; there are 4 of them in total, and each has a single trace in it.

Now, let us assume the program $a_1a_2 \parallel b_1b_2$ is not disjoint parallel anymore and $a_1$ and $b_1$ are right movers. This means that $P$ can still be reduced to $A \parallel B$ with the same atomic blocks. Note that Lipton's definition of a right mover is strictly stronger than (and therefore implies) our definition of semi-commutativity. That is, if $a_1$ is a right mover, then for all $l \in \stmt$, $(a_1, l) \in I$. Therefore, we have a sound semi-independence relation
\[I = \{(a_1,b_1),(a_1,b_2),(b_1,a_1),(b_1,a_2)\}.\]

\begin{wrapfigure}{r}{0.4\textwidth}
\vspace{-15pt}
\begin{center}
\includegraphics[scale=0.2]{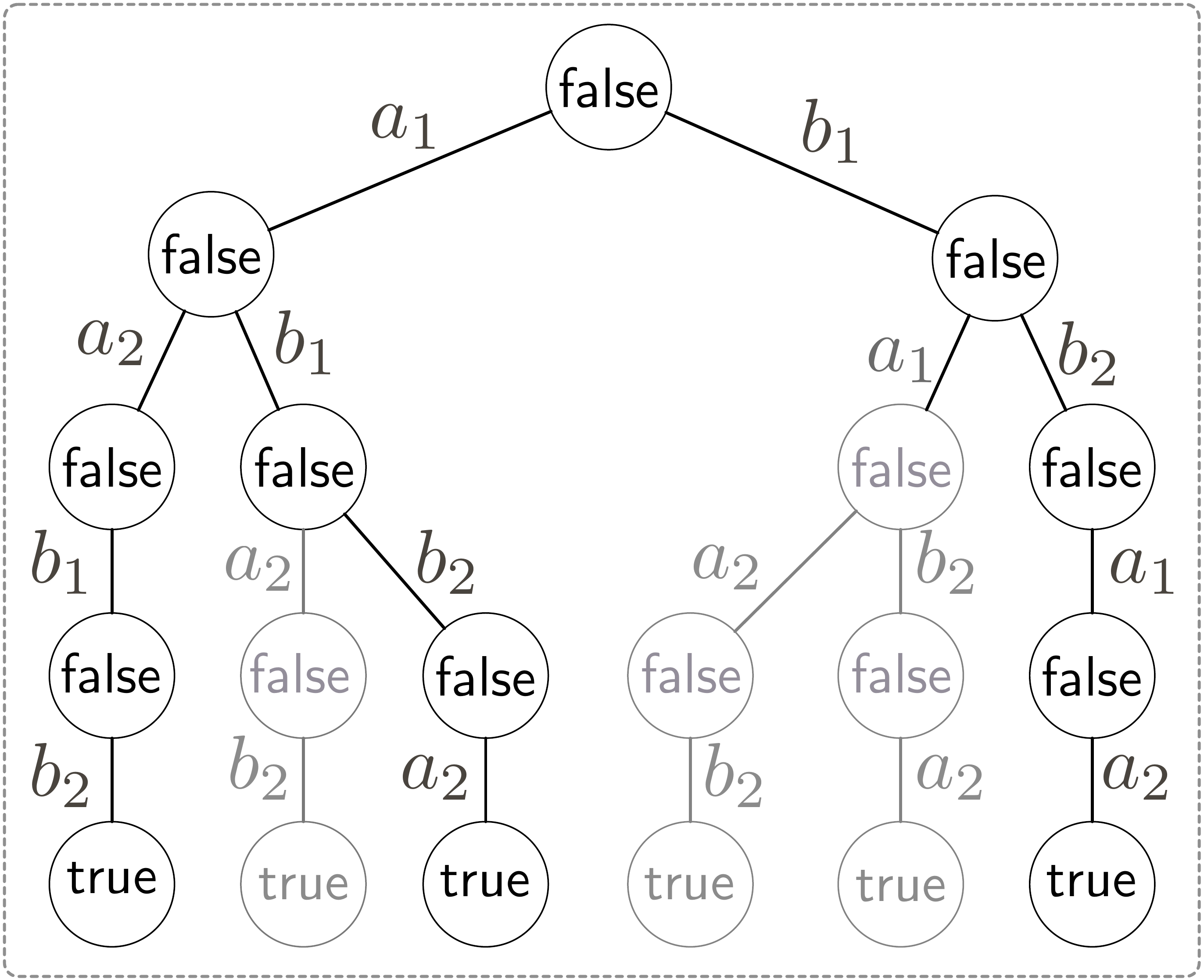}\vspace{-10pt}
\caption{S-reductions vs Lipton's reductions.} \label{fig:lip}
\end{center}
\vspace{-15pt}
\end{wrapfigure}
The program is upwards-closed with respect to $I$. One of the S-reductions of the program is illustrated in Figure \ref{fig:lip}.  Assume that at every relevant node, the $a$ transition is explored first (which would correspond to the prefix order traversal of the depicted tree). The figure illustrates which runs are pruned (greyed out) and which remain in the reduction. Specifically, the {\em unexpected} trace $a_1 b_1 b_2 a_2$
has to remain in the reduction because the trace $b_1 b_2 a_1 a_2$ which subsumes it is only visited {\em later}, and cannot be used for pruning. Therefore, this particular s-reduction has an extra trace compared to Lipton's reduction $A \parallel B$. The rest of the choices of order (beyond always exploring a's before b's at every point) will follow a similar pattern, and include other redundant runs. It is easy to check that all four different S-reductions of this program include redundant traces in addition to Lipton's reduction.

The combination of the two examples demonstrate how our S-reductions and Lipton's (non-contextual) reductions are incomparable and therefore complementary. Note that for any concurrent program (with a bounded number of threads), there are only finitely many choices for atomic blocks. Therefore, one can imagine enumerating them all as a specialized reduction class. Since there are finitely many possible reductions, the class of reductions is trivially recognizable by an LTA. However, our generic C-reduction (or S-reduction) classes do not necessarily include all these (finitely many) reductions.

%

\subsection{Existential Boundedness}

\begin{wrapfigure}{r}{0.35\textwidth}\vspace{-20pt}
\begin{center}
\includegraphics[scale=0.28]{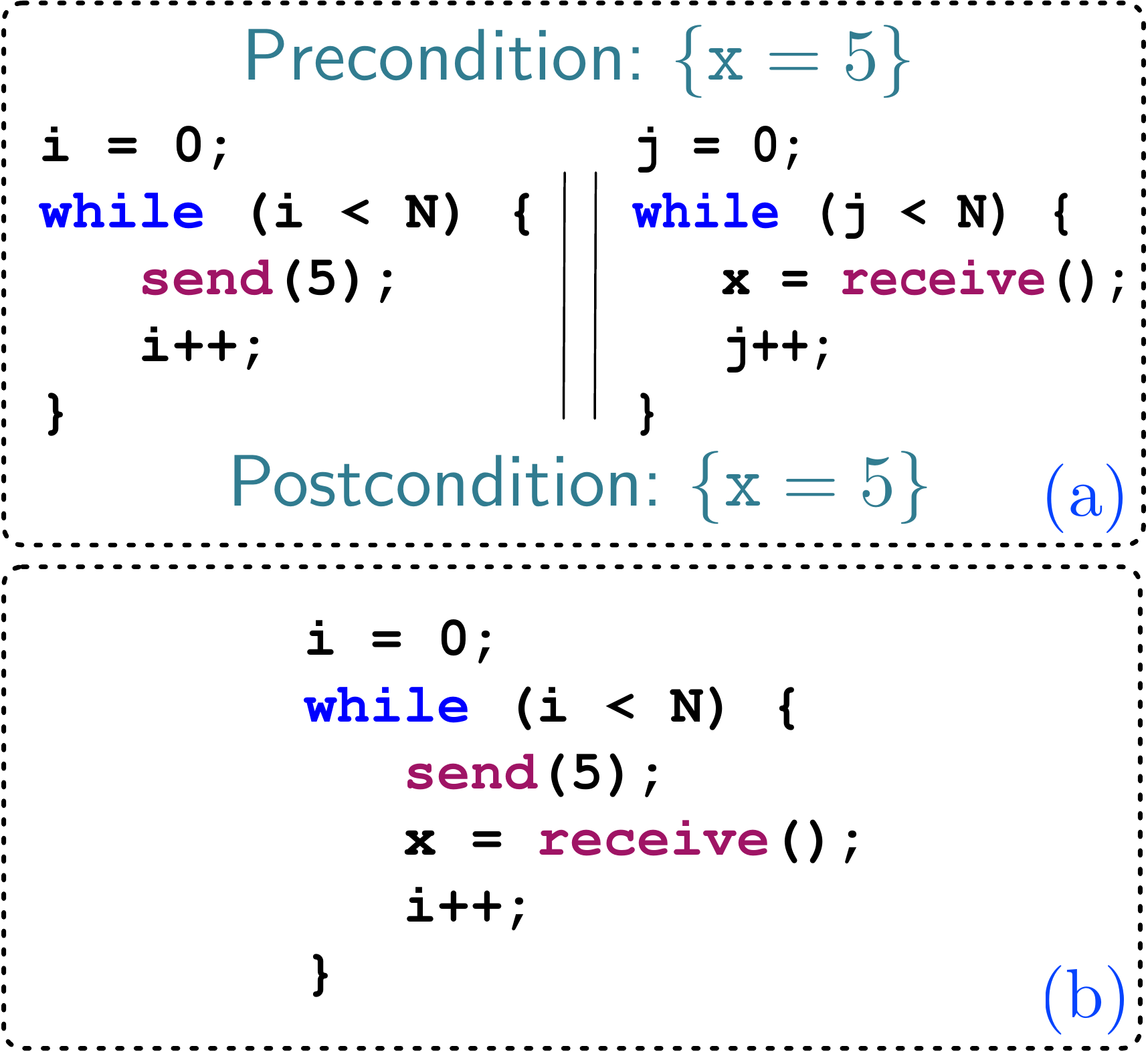}\vspace{-10pt}
\caption{A Synchronous Reduction.} \label{fig:eb1}\vspace{-5pt}
\end{center}\vspace{-10pt}
\end{wrapfigure}

Programs operating on unbounded FIFO channels typically require quantified invariants for their correctness proofs. For example, the simple code in Figure \ref{fig:eb1}(a)   requires an invariant stating that all elements in transit at any given time are equal to 5. Note, however, that every trace of this program is equivalent to a trace of the program in Figure \ref{fig:eb1}(b) where every receive occurs right after its corresponding send. For each trace in Figure \ref{fig:eb1}(b), there is at most one element in transit at any given time, which makes the program effectively finite state. The proposed approaches in \cite{Desai14,psynch}, for example, verify the program in Figure \ref{fig:eb1}(a) precisely by transforming it to the one in Figure \ref{fig:eb1}(b).

The program in Figure \ref{fig:eb1}(b) is called \emph{universally bounded}, since there is a fixed bound $k$ ($= 1$ in this case) where no execution of the program has more than $k$ messages in transit at any given time. Conversely, the program in Figure \ref{fig:eb1}(a) is called \emph{existentially bounded}, because every execution of this program is equivalent to some execution where no more than $k$ messages are in transit at any given time. Universally bounded programs are usually much easier to verify because their channels are bounded. We argue that if a program is existentially bounded then there exists a universally bounded C-reduction.
Note that our algorithm does not have to be aware of the existential boundedness of its input. The claim is that if the input is existentially bounded, the C-reductions will provide the opportunity for the simpler proof to be found.

To provide the formal result, we use the setup similar to the one in \cite{Genest07} (in this section only). We fix a finite set $\mathrm{Chan}$ of unbounded FIFO channels. For simplicity, we assume programs can \emph{only} perform send and receive actions on channels\anthony{TODO: drop this assumption}. Furthermore, we do not concern ourselves with the particular contents of the channels; \azadeh{What does this mean?} we only care about the number of messages in transit at any given time. Formally, we instantiate our state set to $\state = (\mathrm{Chan} \to \mathbb{N})$ and our alphabet to $\stmt = \{ \mathrm{send}(c), \mathrm{recv}(c) \mid c \in \mathrm{Chan} \}$. We assign the semantics
\begin{align*}
  \sem{\mathrm{send}(c)} &= \{ (f, f[c \mapsto f(c) + 1]) \mid f \in \state\} \\
  \sem{\mathrm{recv}(c)} &= \{ (f, f[c \mapsto f(c) - 1]) \mid f \in \state, f(c) > 0 \}
\end{align*}
where $f[x \to y]$ denotes function $f$ with the output for $x$ replaced with $y$.

A trace $\tau$ is said to be \emph{$k$-bounded} if the number of elements in transit in any given channel never exceeds $k$. Formally, this means that for any prefix $\sigma$ of $\tau$, we have
\[
  (\lambda \_ \ldotp 0, f) \in \sem{\sigma} \implies f(c) \le k
\]
for all $f \in \state$ and $c \in \mathrm{Chan}$. We say a program $P$ is \emph{existentially $k$-bounded} if every trace of $P$ is equivalent (with respect to the symmetric subset of $\sqsubseteq_{\mathcal{I}^\mathrm{sound}_P}$, which we shall denote by $\sim_{\mathcal{I}^\mathrm{sound}_P}$) by a $k$-bounded trace. We say $P$ is \emph{universally $k$-bounded} if every trace of $P$ is $k$-bounded. With these definitions in place, we present the main theorem:
\begin{theoremrep}\label{thm:eb}
  If $P$ is an existentially $k$-bounded program, then there exists a universally $k$-bounded reduction $\red{P}{\mathcal{I}_P^\mathrm{sound}}{O} \in \creduce_{\mathcal{I}_P^\mathrm{sound}}(P)$ for some exploration ordering $O : \stmt^* \to \linear{\stmt}$.
\end{theoremrep}
\begin{proof}
  We write $\mathrm{chan}(a)$ to denote the channel on which a statement operates, that is
  \begin{align*}
    \mathrm{chan}(\mathrm{send}(c)) &= c \\
    \mathrm{chan}(\mathrm{recv}(c)) &= c
  \end{align*}

  Define $O(\sigma)$ to return some ordering $\le_\stmt$ that explores statements in an order that minimizes buffer sizes. For example, if buffer $c$ has 4 elements after executing $\sigma$ and buffer $c'$ has 5, then $\mathrm{send}(c)$ is explored before $\mathrm{send}(c')$. In all cases, receives are explored before sends.

  Assume $x \in P\downarrow_{\mathcal{I}_P^\mathrm{sound}, O}$ and $y$ is some $k$-bounded trace such that $x \sim_{\mathcal{I}_P^\mathrm{sound}} y$. It suffices to show that $x$ is also $k$-bounded.

  We proceed by (indexed) lexicographical induction on $y$, using $<_O$ defined in the proof of Lemma \ref{lem:semi-pruned}. Our inductive hypothesis is: if $y'$ is some $k$-bounded trace such that $x \sim_{\mathcal{I}_P^\mathrm{sound}} y'$ and $y'$ is lexicographically smaller than $y$, then $x$ is $k$-bounded.

  If $x = y$, then we’re done, so assume $x \ne y$. Then $x$ and $y$ must have some common prefix $z$ followed by differing statements $a$ and $b$ such that

  \begin{itemize}
    \item $x = zax_1bx_2$, and
    \item $y = zby_1ay_2$
  \end{itemize}

  for some traces $x_1, x_2, y_1, y_2$ such that

  \begin{itemize}
    \item $b$ is independent of $ax_1$ in context $z$ (i.e. $z a x_1 b \sim_{\mathcal{I}_P^\mathrm{sound}} z b a x_1$), and
    \item $a$ is independent of $by_1$ in context $z$ (i.e. $z b y_1 a \sim_{\mathcal{I}_P^\mathrm{sound}} z a b y_1$).
  \end{itemize}

  Since $x \in P\downarrow_{\mathcal{I}_P^\mathrm{sound}, O}$ the above implies that $a$ is explored before $b$ (i.e. $(b, a) \in O(z)$), or else $x$ would be pruned. Let $y' = zaby_1y_2$ (i.e. $y'$ is the trace obtained by moving $a$ to the left of $by_1$). Then $y' \sim_{\mathcal{I}_P^\mathrm{sound}} y \sim_{\mathcal{I}_P^\mathrm{sound}} x$ and $y'$ is lexicographically smaller than $y$. By the inductive hypothesis, it suffices to show that $y'$ is $k$-bounded.

  We proceed by cases:

  \begin{itemize}
  \item Assume $a = \mathrm{recv}(q)$ for some queue $q$. Moving $\mathrm{recv}$ statements to the left never increases $k$-boundedness, so $y'$ is $k$-bounded.
  \item Assume $a = \mathrm{send}(q)$. A statement cannot commute with itself, so every statement in $y_1$ must either be $\mathrm{recv}(q)$ or some other operation not involving $q$. Let $n$ be the number of $\mathrm{recv}(q)$ statements in $by_1$, such that $by_1 = y_{10} \cdot \mathrm{recv}(q) \cdot \ldots \cdot \mathrm{recv}(q) \cdot y_{1n}$ where $\mathrm{send}(q), \mathrm{recv}(q) \notin y_{1i}$ for each $0 \le i \le n$.
    \begin{itemize}
      \item Assume $n = 0$. Moving $\mathrm{send}$ statements to the left of other operations on different queues does not affect $k$-boundedness, so $y'$ is $k$-bounded.
      \item Assume $z$ is infeasible. Then moving $a$ to the left of $by_1$ cannot affect $k$-boundedness since $a$ never appears in a feasible position, so $y'$ is $k$-bounded.
      \item Assume $n > 0$ and $z$ is feasible. We can move $a$ to the left of each $\mathrm{recv}(q)$ statement in $by_1$ except for the first without affecting $k$-boundedness (i.e. $zby_{10} \cdot \mathrm{recv}(q) \cdot \mathrm{send}(q) \cdot \ldots \cdot \mathrm{recv}(q) \cdot y_{1n}y_2$ is $k$-bounded). We can show this in two cases:
      \begin{itemize}
          \item If the first $\mathrm{recv}(q)$ is feasible, then $q$ must have some number $m + 1$ elements in it after executing $zy_{10}$, so $k \ge m + 1$. After the first $\mathrm{recv}(q)$, $q$ will have $m$ elements, and after executing $\mathrm{send}(q)$, $q$ will once again have $m + 1$ elements. Thus moving the $\mathrm{send}(q)$ does not affect the maximum number of elements that appear in $q$, which preserves $k$-boundedness.
          \item If the first $\mathrm{recv}(q)$ is infeasible, then $a$ never appears in a feasible position even if we move it to the left of all the other $\mathrm{recv}(q)$s, so $k$-boundedness is still preserved.
          Our final task is to show we can push $b$ past the first $\mathrm{recv}(q)$ without affecting $k$-boundedness.

          Since $(b, a) \in O(z)$, it cannot be the case that $y_{10} = \epsilon$ since that would imply $b = \mathrm{recv}(q)$ and therefore $(\mathrm{recv}(q), \mathrm{send}(q)) \in O(z)$, which is a contradiction because $\mathrm{recv}$ statements are always explored before $\mathrm{send}(q)$ statements. Thus $y_{10}$ is non-empty and $b \in \{\mathrm{send}(q'), \mathrm{recv}(q')\}$ for some $q' \ne q$. Once again, $b$ cannot be a $\mathrm{recv}$ statement, so $b = \mathrm{send}(q')$.

          Since $(\mathrm{send}(q'), \mathrm{send}(q)) \in O(z)$, $q$ must have at most as many elements as $q'$ after executing $z$. If $\mathrm{send}(q)$ is executed first, then the maximum number of elements that appear in either of the queues will still be the same, and therefore $y'$ is $k$-bounded.

        \end{itemize}
    \end{itemize}
  \end{itemize}
\end{proof}
There are existing methods \cite{Desai14,psynch}, designed specifically to transform asynchronous programs into synchronous ones. The significance of Theorem \ref{thm:eb} is that it demonstrates that even though our proposed technique is not specialized for this category, it can potentially behave well on existentially bounded programs.

It should be noted that our technique cannot effectively use Theorem \ref{thm:eb} unless it is also able to come up with a soundness proof for $\mathcal{I}_P^\mathrm{sound}$. This is not always possible because such proofs would generally have to include assertions that somehow ``count'' the number of elements in each queue, yielding a non-regular language. However, we conjecture that a weaker independence relation will always suffice.
\begin{conjecture}
  If $P$ is an existentially $k$-bounded program, then there exists a universally $k$-bounded reduction $\red{P}{\mathcal{I}}{O} \in \creduce^*(P)$ for some independence relation $\mathcal{I} : \stmt^* \to \pow{\stmt \times \stmt}$ and exploration ordering $O : \stmt^* \to \linear{\stmt}$ such that there exists a simple proof $\Pi$ (i.e. $\Pi$ is in linear integer arithmetic) such that $\mathrm{sound}(\mathcal{I}) \subseteq \lang{\Pi}$.
\end{conjecture}



\section{Refinement-style Verification Algorithm}\label{sec:algorithm}

\begin{figure}[htbp]
\begin{center}\vspace{-15pt}
\includegraphics[scale=0.2]{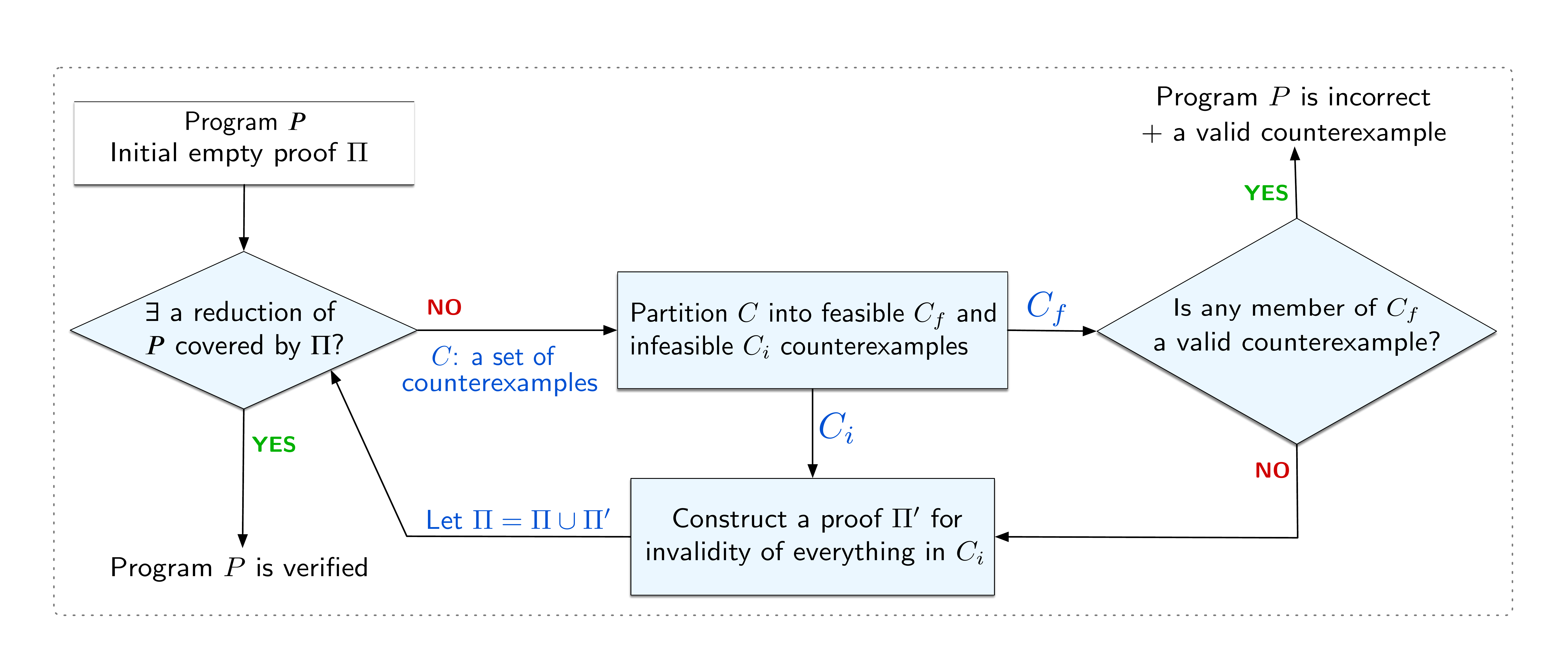}\vspace{-15pt}
\caption{Counterexample-guided refinement loop.}\vspace{-10pt}
\label{fig:loop}
\end{center}
\end{figure}

Figure \ref{fig:loop} illustrates the outline of our verification algorithm. It is a counterexample-guided abstraction refinement loop in the style of \cite{FarzanKP13,FarzanKP15,HeizmannHP09,cav19}. The algorithm starts with assertions {\sf true} and {\sf false} in the proof and iteratively discovers more assertions until a proof is found. Unlike standard refinement loops, it suffices to discover the proof for a {\em sound} reduction of the program and not the entire program.

To check the validity of any candidate proof $\Pi$, one has to check if there exists a reduction of the program that is covered by $\Pi$. The results presented in this paper so far naturally give rise to an algorithm for performing this check:
\begin{itemize}
\item The set of program reductions are LTA-representable (Theorems \ref{thm:reduce-lta} and \ref{thm:reduce-lta-2}).
\item A regular proof can be constructed based on a candidate set of assertions (see Section \ref{sec:ptrace}).
\item LTA representability of the program and regularity of the proof language imply that the check can be performed through an emptiness test for the intersection of LTA languages (Theorem \ref{thm:pc-decidable}). There are known algorithms for both tasks \cite{BaaderT01}.
\end{itemize}

When the proof is not sufficient, a {\em finite} set of counterexamples can be produced as a witness. These counterexamples are sequences of statements. One can check whether they are feasible or not. In a standard setting, a feasible counterexample would be a witness for the violation of the property by the program. In our setting, not every counterexample is a program trace. If a program trace counterexample is feasible, then the program is unsafe. We stop the loop and return the counterexample.

The specific construction of our LTA for C-reductions implies that some reductions in the class may be unsound. Any feasible counterexample that is not a program trace is a witness for the unsoundness of (at least) one such reduction. We refer to these as {\em invalid} counterexamples. These are therefore spurious counterexamples and are ignored. We address how this does not affect the termination of the refinement loop in Section \ref{sec:progress}.

If no true counterexample is found, then one can produce proofs of infeasibility of all the infeasible counterexamples with the aid of any program prover. All new assertions discovered through this process are then added to the current proof conjecture, and the refinement loop restarts. Note that proofs of infeasibility of program trace counterexamples contribute towards the discovery of a program proof, and proofs of infeasibility of the rest would contribute towards discovery of invariants that expand the set of sound contextual commutativity relations.
In our tool, we use Craig interpolation to produce proofs of infeasibility of these counterexamples. In general, since program traces are the simplest forms of sequential programs (loop and branch free), any automated program prover (that can handle proving them) may be used.

\subsection{Termination, Soundness and Completeness}\label{sec:progress}

Let us assume that the program is correct, and more specifically, there exists a proof $\Pi^*$ that subsumes one of its contextual reductions in $\creduce^*(P)$. Ideally, we would like to claim that the refinement algorithm of  Figure \ref{fig:loop} succeeds under these conditions.

Note that the convergence of the algorithm depends on two factors: (1) the counterexamples used by the algorithm belong to $\lang{\Pi^*}$ and (2) the proofs discovered by the backend solver for these counterexamples use assertions from $\Pi^*$. The latter is a typical known wild card in software model checking, which cannot be guaranteed; there is plenty of empirical evidence, however, that procedures based on Craig Interpolation do well in approximating it. This problem is orthogonal to the contribution of this paper.

The former poses a new problem specific to our methodology: if the ideal proof $\Pi^*$ is the target of the refinement loop, since the set of traces proved correct in $\lang{\Pi^*}$ is incomparable to the set of program traces $P$, then one cannot just use any program trace as an \emph{appropriate} counterexample.


A key observation from LTA's can be used to solve this problem. For an LTA $M$ and regular language $L$, when there exists no $R \in \lang{M}$ such that $R \subseteq L$, then there exists a finite set of counterexamples $C$ such that for all $R \in \lang{M}$, there exists some $\tau \in C$ such that $\tau \in R$ and $\tau \not \in L$ \cite{cav19}. This is very significant, since it means that when we check a proof candidate $\lang{\Pi}$, and it does not cover any program reduction, then there are finitely many counterexamples that together {\em cover} all reductions. This means that if all these counterexamples are proved infeasible, and $\Pi$ is updated with these proofs, then the proof has made meaningful progress for every single reduction.

But what about invalid counterexamples from the set $C_f$ (of Figure \ref{fig:loop}) which our algorithm simply ignores? Note that these may appear again in subsequent rounds. Fortunately, this behaviour turns out to be unproblematic: strong progress essentially relies on finding new counterexamples for each \emph{correct} reduction. Whether we find new (or even any) counterexamples for incorrect reductions is of no importance.

\begin{theoremrep}[Strong Progress]\label{thm:stp}
   Assume there exists a reduction $P^* \in \creduce^*(P)$ (or alternatively $\reduce^*(P)$) and a set of assertions $\Pi^*$, such that $P^* \subseteq \lang{\Pi^*}$. If the algorithm of Figure \ref{fig:loop} uses assertions from $\Pi^*$ to prove the infeasibility of those counterexamples which belong to $\lang{\Pi^*}$, then it will terminate in at most $|\Pi^*|$ iterations.
\end{theoremrep}
\begin{proof}
   It is sufficient to show that we learn at least one new assertion in $\Pi^*$ every iteration. Assume we have received a counterexample set $C$ such that, for all $P' \in \creduce^*(P)$, there exists some $x \in C$ such that $x \in P'$ and $x \notin \lang{\Pi}$ (\cite{cav19} ensures $C$ exists). Let $x^* \in C$ be the counterexample for $P^*$. Then \textsc{Interpolate}$(x)$ will return new assertions $\Pi' \subseteq \Pi^*$ satisfying $x^* \in \lang{\Pi'}$. If $\Pi' \subseteq \Pi$ then $x^*$ would not have been returned as a counterexample, so there must exist some $\phi \in \Pi'$ (and therefore $\phi \in \Pi^*$) such that $\phi \notin \Pi$.
\end{proof}

Theorem \ref{thm:stp} ensures that the algorithm will never get into an infinite loop due to a bad choice of counterexamples. The extra condition on proofs of traces from $\Pi^*$ rules out diverging behaviour that could occur due to the wrong choice of assertions by the backend prover. A wrong choice of assertions can cause divergence in any standard software model checking algorithm (even for sequential integer programs with simple proofs) that relies on discovery of loop invariants through interpolation. The assumption that there exists a proof for a reduction (in the fixed set $\creduce^*(P)$) of the program ensures that the algorithm is searching for a proof that does exist. Note that, in general, a proof may exist for a reduction of the program which is not in $\creduce^*(P)$. That is, the algorithm is not complete with respect to all reductions, but only reductions in $\creduce^*(P)$. Since checking the premises of \ref{rule:safety+reductions-bad} for all semantic reductions is undecidable as discussed in Section \ref{sec:background}, a complete algorithm does not exist. The soundness of the algorithm is a straightforward consequence of the soundness of reductions stated in Sections \ref{sec:semi} and \ref{sec:context}.


\section{An Efficient Algorithm for Proof Checking}\label{sec:optimization}

\begin{toappendix}
\label{apx:optimization}
\end{toappendix}

The proof checking algorithm described in Section \ref{sec:algorithm} boils down to an emptiness check on the intersection (i.e. product) of the LTAs representing all program reductions and the proof language. The size of the reduction LTA is exponential on the input program size, both in terms of the number of states and the number of transitions per state. This can make proof checking prohibitively expensive, even though the emptiness test is performed in linear time. In this section, we propose a new algorithm for proof checking that has a provably better worst-case time complexity and works better in practice. First, we show how to drastically reduce the number of transitions considered during proof checking. This also allows us to easily employ antichain-based optimizations (in the style of \cite{WulfDHR06,cav19}) to better deal with the exponential state space, which in turn allows us to further reduce the number of transitions considered and arrive at an asymptotically better algorithm than the one given in \cite{cav19}.

\subsection{Proof-Driven Maximal Independence Relations}\label{sec:maximal}

Our construction of the reduction LTA enumerates a class of contextual independence relations and a class of reductions that are induced by them. The key observation of this section is that a specific proof candidate instigates some additional structure in the state space of all contextual independence relations that would allow us to {\em soundly and completely} choose one {\em maximal} relation instead of exploring them all.


For the purpose of checking a proof candidate $\Pi$, it suffices to only consider those independence relations that are correct according to $\Pi$, i.e. all independence relations $\mathcal{I} : \stmt^* \to \pow{\stmt \times \stmt}$ such that $\mathrm{sound}(\mathcal{I}) \subseteq \lang{\Pi}$. The proof checking algorithm will never certify a reduction based on an unproven independence relation anyways. In fact, even among independence relations that are correct according to $\Pi$, it suffices to consider only a single \emph{maximal independence relation} $\mathcal{I}_\Pi \subseteq \mathcal{I}_P$ induced by the proof $\Pi$, defined as
\[
  \mathcal{I}_\Pi(\sigma) = \{ (a, b) \in \mathcal{I}_P(\sigma) \mid \sigma \cdot \mathrm{indep}_{a, b} \in \lang{\Pi} \}.
\]
This independence relation declares a pair of statements independent precisely when it is sound to do so according to $\Pi$. By maximal, we mean that any independence relation $\mathcal{I}$ that is sound according to $\Pi$ is subsumed by $\mathcal{I}_\Pi$.
\begin{proposition}\label{prop:ipi}
  For any independence relation $\mathcal{I} \subseteq \mathcal{I}_P$, if $\mathrm{sound}(\mathcal{I}) \subseteq \lang{\Pi}$ then $\mathcal{I} \subseteq \mathcal{I}_\Pi$.
\end{proposition}
$\mathcal{I}_\Pi$ can be shown to be regular by modifying the automaton recognizing $\lang{\Pi}$. By Theorem \ref{thm:reduce-lta-2}, $\creduce_{\mathcal{I}_\Pi}(P)$ is representable by an LTA. The following theorem states that proof checking against $\mathcal{I}_\Pi$ is just as good as proof checking against all independence relations.
\begin{theoremrep}
  There exists some $P' \in \creduce^*(P)$ satisfying $P' \subseteq \lang{\Pi}$ iff there exists some $P'' \in \creduce_{\mathcal{I}_\Pi}(P)$ satisfying $P'' \subseteq \lang{\Pi}$.
\end{theoremrep}
\begin{proof}
  $(\rightarrow)$ By the definition of $\creduce^*(P)$, we have $P' = \red{P}{\mathcal{I}}{O} \cup \mathrm{sound}(\mathcal{I})$ for some $\mathcal{I} \subseteq \mathcal{I}_P$ and $O : \stmt^* \to \linear{\stmt}$. Thus $\red{P}{\mathcal{I}}{O} \subseteq \lang{\Pi}$ and $\mathrm{sound}(\mathcal{I}) \subseteq \lang{\Pi}$. By Proposition \ref{prop:ipi}, we have $\mathcal{I} \subseteq \mathcal{I}_\Pi$. Using similar reasoning to the proof of Proposition \ref{prop:red-subsumption}, we may conclude $\red{P}{\mathcal{I}_\Pi}{O} \subseteq \red{P}{\mathcal{I}}{O}$. Therefore this case is satisfied for $P'' = \red{P}{\mathcal{I}_\Pi}{O}$.

  $(\leftarrow)$ By the definition of $\creduce_{\mathcal{I}_\Pi}(P)$ we have $P'' = \red{P}{\mathcal{I}_\Pi}{O}$ for some $O : \stmt^* \to \linear{\stmt}$. By the definition of $\mathcal{I}_\Pi$ we have $\mathrm{sound}(\mathcal{I}_\Pi) \subseteq \lang{\Pi}$, so this case is satisfied for $P' = \red{P}{\mathcal{I}_\Pi}{O}$.
\end{proof}
Therefore, we can replace the LTA representing $\creduce^*(P)$ with the LTA representing $\creduce_{\mathcal{I}_\Pi}(P)$ in our proof checking algorithm. While this reduces the number of transitions considered exponentially, it also increases the state space of the reduction LTA since $\creduce_{\mathcal{I}_\Pi}(P)$ must simulate the automaton witnessing regularity of $\mathcal{I}_\Pi$. This turns out to be of no consequence: the proof automaton already contains the state space of $\mathcal{I}_\Pi$, so the state space of the product automaton remains unchanged. Thus we obtain a product automaton with exponentially fewer transitions and an identical state space, resulting in an exponentially faster proof checking algorithm.

\subsection{Optimizing Emptiness Test Through Antichains}\label{sec:antichains}

While LTAs can be checked for emptiness in linear time, the size of the checked automaton is exponential in $|\stmt|$ (in the worst case), since the reduction LTA (as described in Section \ref{sec:repsem}) must maintain \sset sets. In \cite{cav19}, this problem is alleviated using antichain methods \cite{WulfDHR06} for the case of symmetric and non-contextual independence relations. The idea is that emptiness checking reduces to constructing a set of \emph{inactive} states from which no language is accepted. The set of inactive states is shown to be downwards-closed with respect to a particular subsumption relation, which allows the set to be represented compactly by its maximal elements (i.e. antichains).

It turns out that with the contextual (and semi-) independence relations, we can also adapt these methods as an optimization for proof checking. This comes as a positive byproduct of the use of the maximal independence relation $\mathcal{I}_\Pi$ that we described in Section \ref{sec:maximal}. The key observation is that we can pretend that we are in the non-contextual setting of \cite{cav19}, but recover all the relevant information about $\mathcal{I}_\Pi$ from the state of the product automaton that contains information about both program and proof automaton states. Therefore, we can recover the required information about $\mathcal{I}_\Pi$ and know what transitions are independent at each state of the product automaton.

\begin{toappendix}
  \subsection*{Description of Contextual Antichain Algorithm}
  Assume the program $P$ is represented by the automaton $A_P = (Q_P, \stmt, \delta_P, q_{0P}, F_P)$, and assume the proof language $\lang{\Pi}$ is represented by the automaton $A_\Pi = (Q_\Pi, \stmt, \delta_\Pi, q_{0\Pi}, F_\Pi)$. We recall the definition of $F^{\max} : (Q_P \times Q_\Pi \to \pow{\pow{\stmt}}) \to (Q_P \times Q_\Pi \to \pow{\pow{\stmt}})$ from \cite{cav19}.
  \[
    F^{\max}_{M_{P\Pi}}(X)(q_P, q_\Pi) = \begin{cases}
      \{ \stmt \} & \text{if $q_P \in F_P \land q_\Pi \notin F_\Pi$} \\
      \bigsqcap \limits_{O \in \linear{\stmt}}
      \bigsqcup \limits_{\substack{
        a \in \Sigma \\
        S \in X(q'_P, q'_\Pi)}}
        S'
      & \text{otherwise}
  \end{cases}
  \]
  where
  \begin{align*}
    q'_P &= \delta_P(q_P, a) &
    X \sqcap Y &= \operatorname{max} \{ x \cap y \mid x \in X \land y \in Y \} \\
    q'_\Pi &= \delta_\Pi(q_\Pi, a) &
    X \sqcup Y &= \operatorname{max} (X \cup Y)
  \end{align*}
  \[
    S' = \begin{cases}
      \{ (S \cup \overline{I(a)}) \setminus \{a \} \} & \text{if $O(a) \cap I(a) \subseteq S$} \\
      \emptyset & \text{otherwise}
    \end{cases}.
  \]
  and $I$ is the static (symmetric) independence relation. This definition needs no modifications to support semi-independence; our definitions of $\sleep$ and $\ignore$ given in the proofs in Section \ref{sec:semi} are identical to the ones given in \cite{cav19}.

  Let $A_{\mathcal{I}_P} = (Q_P, \stmt, \delta_P, q_{0P}, F_{\mathcal{I}_P})$ be the automaton witnessing regularity of $\mathcal{I}_P$ (recall from Section \ref{sec:context} that this automaton is identical to $A_P$ aside from its final ``states''). To accommodate contextuality, we need only replace $I$ in the above definition with the relation
  \[
    I' = \{ (a, b) \in F_{\mathcal{I}_P}(q_P) \mid \delta(q_\Pi, \mathrm{indep}_{a, b}) \in F_\Pi \}
  \]
  which consists of all pairs of statements where $P$ is closed at state $q_P$ ($(a, b) \in F_{\mathcal{I}_P}(q_P)$) and soundly commute according to $\Pi$ at state $q_\Pi$ ($\delta(q_\Pi, \mathrm{indep}_{a, b}) \in F_\Pi$).

  Intuitively, we want to replace $I$ with $\mathcal{I}_\Pi(\sigma)$, for some appropriate $\sigma$ such that $\delta_\Pi(q_{0\Pi}, \sigma) = q_\Pi$ and $\delta_P(q_{0P}, \sigma) = q_P$. Then we have
  \begin{align*}
    \mathcal{I}_\Pi(\sigma)
    &= \{ (a, b) \in \mathcal{I}_P(\sigma) \mid \sigma \cdot \mathrm{indep}_{a, b} \in \lang{\Pi} \} \\
    &= \{ (a, b) \in \mathcal{I}_P(\sigma) \mid \delta^*_\Pi(q_{0\Pi}, \sigma \cdot \mathrm{indep}_{a, b}) \in F_\Pi \} \\
    &= \{ (a, b) \in \mathcal{I}_P(\sigma) \mid \delta_\Pi(\delta^*(q_{0\Pi}, \sigma), \mathrm{indep}_{a, b}) \in F_\Pi \} \\
    &= \{ (a, b) \in F_{\mathcal{I}_P}(\delta^*_P(q_{0P}, \sigma)) \mid \delta_\Pi(q_\Pi, \mathrm{indep}_{a, b}) \in F_\Pi \} \\
    &= \{ (a, b) \in F_{\mathcal{I}_P}(q_P) \mid \delta_\Pi(q_\Pi, \mathrm{indep}_{a, b}) \in F_\Pi \} \\
    &= I'.
  \end{align*}

  As in \cite{cav19}, proof checking passes iff the least fixed-point of $F^{\max}$ is empty at states $(q_{0P}, q_{0\Pi})$, i.e. $\mathrm{lfp}(F^{\max})(q_{0P}, q_{0\Pi}) = \emptyset$.
\end{toappendix}

Antichain methods do not generally improve the worst-case computational complexity of an algorithm, since the size of the largest antichain on sets of a finite alphabet is still exponential in the size of the alphabet. However, antichains are never larger than the sets they represent, and often exponentially smaller, making antichain-based algorithms very efficient in practice.

\subsection{Time Complexity of Proof Checking}

The final source of complexity that we would like to eliminate is a factorial component that arises because our reduction automaton considers all possible exploration strategies of the program (through enumerations of all order functions). In particular, each state in the reduction automaton has a transition for all $|\stmt|!$ linear orderings over $\stmt$. In conjunction with the optimization of Section \ref{sec:antichains}, this translates to an iterated antichain meet over all linear orders, which has exponential-of-factorial complexity. Fortunately, we can reduce this factor to only exponential in the size of $\stmt$.

As mentioned previously, LTA emptiness is calculated via a fixed point computation. The complexity occurs within the function over which the fixed point is computed. Therefore, we shall focus on the complexity of calculating this function. This function, which we call $F^{\max}$, has type
\[
  F^{\max} : (Q_P \times Q_\Pi \to \pow{\pow{\stmt}}) \to (Q_P \times Q_\Pi \to \pow{\pow{\stmt}}),
\]
where $Q_P$ and $Q_\Pi$ are respectively the state spaces of the DFAs accepting the program language $P$ and the proof language $\lang{\Pi}$. Since the input of $F^{\max}$ is itself a function, we define the size of a function $X : Q_P \times Q_\Pi \to \pow{\pow{\stmt}}$ to be the maximum size of all its outputs, i.e.
\[
  |X| = \max \{ |X(q_P, q_\Pi)| \mid q_P \in Q_P, q_\Pi \in Q_\Pi \}.
\]

\begin{theoremrep}
  The algorithm as of Section \ref{sec:antichains} computes $F^{\max}(X)$ in $\mathcal{O}((|\stmt||X|)^{2|\stmt|!})$ time.
\end{theoremrep}
\begin{proof}
  The definition of $F^{\max}$ given above involves an iterated antichain meet of an iterated antichain join. Given two arguments $A$ and $B$, antichain meets are calculated in $\mathcal{O}((|A||B|)^2)$ time by finding the maximal elements amongst the pairwise intersections of $A$ and $B$. Conversely, antichain joins are calculated in $\mathcal{O}(|A||B|)$ time by finding the maximal elements among $A \cup B$. An \emph{iterated} antichain meet over $n$ elements of size at-most $m$ has $\mathcal{O}(m^{2n})$ complexity, and an iterated antichain join has $\mathcal{O}(n^2m^2)$ complexity.

  The inner join in $F^{\max}$ given above is over $|\stmt||X|$ antichains of size 1, and therefore has $(|\stmt||X|)^2$ complexity. The outer join is over $|\stmt|!$ antichains of size $\stmt||X|$, and therefore has $\mathcal{O}((|\stmt||X|)^{2|\stmt|!})$. Together, this has $\mathcal{O}((|\stmt||X|)^{2|\stmt|!} + |\stmt|!(|\stmt||X|)^2) = \mathcal{O}((|\stmt||X|)^{2|\stmt|!})$ complexity.
\end{proof}

There is a key observation that allows us to improve these results. The following lemma captures this by effectively permitting the elimination of the set of linear orders from our calculations:
\begin{lemmarep} \label{lem:key} Let $A \subseteq \stmt \times \pow{\stmt}$ be a relation satisfying $\forall (a, S) \in A \ldotp a \in S$. Then
  \[
    (\forall O \in \linear{\stmt} \ldotp \exists (a, S) \in A \ldotp O(a) \subseteq S)
 \  \iff  \   (\exists \emptyset \subset A' \subseteq A \ldotp \forall (a, S) \in A' \ldotp \overline{\mathrm{Dom}(A')} \subseteq S) \]
  where $\mathrm{Dom}(A')$ is the domain of $A'$.
\end{lemmarep}
\begin{proof}
  $(\rightarrow)$
  Assume $\forall \emptyset \subset A' \subseteq A \ldotp \exists (a, S) \in A' \ldotp \overline{\mathrm{Dom}(A')} \nsubseteq S$.

  We show, by contrapositive, $\exists O \in \linear{\stmt} \ldotp \forall (a, S) \in A \ldotp O(a) \nsubseteq S$.

  Let $B = A$. Then we have $B \subseteq A$. We show
  \[
    \exists O \in \linear{\stmt} \ldotp \forall (a, S) \in B \ldotp O(a) \nsubseteq S
  \]
  by induction on $|B|$ while maintaining $B \subseteq A$.
  \begin{itemize}
    \item If $B = \emptyset$, then the

    \item Assume $B \ne \emptyset$ and $B \subseteq A$.

    For our inductive hypothesis, we assume
    \[
      \forall b \in B \ldotp \exists O \in \linear{\stmt} \ldotp \forall (a, S) \in B \setminus \{b\} \ldotp O(a) \nsubseteq S.
    \]

    We instantiate our first assumption with $A' = B$ to obtain some $(a, S) \in B$ such that $\overline{\mathrm{Dom}(B)} \nsubseteq S$, from which we obtain some $b \notin S$ such that $\forall T \ldotp (b, T) \notin B$.

    By instantiating $b = (a, S)$ in our inductive hypothesis, we obtain some $O \in \linear{\stmt}$ such that
    \[
      \forall (c, U) \in B \setminus \{(a, S)\} \ldotp O(c) \nsubseteq S.
    \]

    Now define $O'$ to be a linear order identical to $O$, except with $b$ at the bottom of the order.

    Then $O'(a) \nsubseteq S$ (since $b \notin O'(a)$ but $b \in S$).

    Also, $\forall (c, U) \in B \setminus \{(a, S)\} \ldotp O'(c) \nsubseteq S$ (since $\forall T \ldotp (b, T) \notin B$, and $b$ is at the bottom of $O'$).

    Thus $\forall (c, U) \in B \ldotp O'(c) \nsubseteq S$.

    Therefore $\exists O \in \linear{\stmt} \ldotp \forall (a, S) \in B \ldotp O(a) \nsubseteq S$, for $O = O'$.
  \end{itemize}

  $(\leftarrow)$
  Assume $\exists \emptyset \subset A' \subseteq A \ldotp \forall (a, S) \in A' \ldotp \overline{\mathrm{Dom}(A')} \subseteq S$.

  Then $\forall (a, S) \in A', b \notin S \ldotp \exists T \ldotp (b, T) \in A'$.

  Since $A' \ne \emptyset$, we have $\mathrm{Dom}(A') \ne \emptyset$.

  Assume $O$ is any linear order on $\stmt$.

  Let $a$ be the maximal element of $\mathrm{Dom}(A')$ with respect to $O$.

  Then there exists some $S$ such that $(a, S) \in A'$.

  Then $(a, S) \in A$ (since $A' \subseteq A$).

  Then $O(a) \subseteq S$. If this were not the case, then there would exist some $b \in O(a)$ such that $b \notin S$, which (by an earlier fact) implies $b \in \mathrm{Dom}(A')$. Necessarily, $a \ne b$, since $(a, S) \in A \implies a \in S$ and $b \notin S$. Since $a$ is maximal, we have $(b, a) \in O$. But $b \in O(a)$, which is a contradiction.

  Thus $\exists (a, S) \in A \ldotp O(a) \subseteq S$.
\end{proof}
The equality implied by the lemma is used to improve the fixed point calculation of $F^{\max}$, based on the LTA built using the construction of Theorem \ref{thm:reduce-lta-2} instantiated by the sound independence relation of Proposition \ref{prop:ipi}. The left-hand side of the equality appears in the fixed point computation, and the right hand side lets us drop the enumeration of all linear orders from it.
This leads us to the following result of reduced overall complexity for the dominant computation part of proof checking:
\begin{theoremrep}\label{thm:opt-complexity}
  $F^{\max}(X)$ can be computed in $\mathcal{O}(2^{|\stmt|}|\stmt||X|)$ time.
\end{theoremrep}
\begin{proof}
  Observe that the formula for $F^{\max}$ given previously ultimately calculates a finite intersection of the form
  \[
    (S_1 \cup \overline{I'(a_1)}) \setminus \{a_1\} \cap \dots \cap (S_n \cup \overline{I'(a_n)}) \setminus \{a_n\}
  \]
  where
  \begin{gather*}
    O(a_i) \subseteq S_i \cup \overline{I'(a_i)} \\
    S_i \in X(\delta_P(q_P, a'), \delta_\Pi(q_\Pi, a'))
  \end{gather*}
  for each $1 \le i \le n$. Define
  \begin{align*}
    \mathcal{P}airs &= \{ (a, S) \mid S \in X(\delta_P(q_P, a), \delta_\Pi(q_\Pi, a)) \} \\
    \mathcal{V}alid &= \{ A \subseteq \mathcal{P}airs \mid \forall O \in \linear{\stmt} \ldotp \exists (a, S) \in A \ldotp O(a) \subseteq S \cap I'(a) \}
  \end{align*}
  Then we can rearrange $F^{\max}$ to
  \[
    \max \left\{ \bigcap\limits_{(a, S) \in A} (S \cup \overline{I'(a)}) \setminus \{a\} \mathrel{\Big|} A \in \mathcal{V}alid \right\}
  \]
  Note that a superset of any $A \in \mathcal{V}alid$ is also in $\mathcal{V}alid$, and adding elements to $A$ will shrink the inner intersection, so we need only minimal elements of $\mathcal{V}alid$.
  \[
    \max \left\{ \bigcap\limits_{(a, S) \in A} (S \cup \overline{I'(a)}) \setminus \{a\} \mathrel{\Big|} A \in \min(\mathcal{V}alid) \right\}
  \]
  By Lemma \ref{lem:key}, we have
  \[
    \mathcal{V}alid = \{ A \subseteq \mathcal{P}airs \mid \exists \emptyset \subset A' \subseteq A \ldotp \forall (a, S) \in A' \ldotp \overline{\mathrm{Dom}(A')} \subseteq S \cup \overline{I'(a)} \}.
  \]
  Consequently, $A \in \min(\mathcal{V}alid) \implies A \ne \emptyset \land \forall (a, S) \in A \ldotp \overline{\mathrm{Dom}(A)} \subseteq S \cup \overline{I'(a)}$. The iteration intersection effectively excludes the domain of $A$, and since each $S \cup \overline{I'(a)}$ is a superset of $\overline{\mathrm{Dom}(A)}$ we can simplify the above to
  \[
    \max \{ \overline{\mathrm{Dom}(A)} \mid A \in \min(\mathcal{V}alid) \}
  \]
  As pointed out earlier, it matters not whether we use $\mathcal{V}alid$ or $\min(\mathcal{V}alid)$. Anything in-between is just as valid.
  \[
    \max \{ \overline{\mathrm{Dom}(A)} \mid \emptyset \subset A \subseteq \mathcal{P}airs \land \forall (a, S) \in A \ldotp \overline{\mathrm{Dom}(A)} \subseteq S \cup \overline{I'(a)} \}
  \]
  Finally, we expand $\mathcal{P}airs$ and simplify to obtain
  \[
    F^{\max}(X)(q_P, q_\Pi) = \max \{ \overline{B} \mid \forall a \in B \ldotp \exists S \in X(\delta_P(q_P, a), \delta_\Pi(q_\Pi, a)) \ldotp \overline{B} \subseteq S \cup \overline{I'(a)} \}
  \]
  This version of $F^{\max}$ can be implemented by iterating over all subsets of $\stmt$, checking whether the condition in the set comprehension holds for each subset, and then taking the maximal elements. Obtaining the correct subsets takes $\mathcal{O}(2^\stmt |\stmt||X|)$ time.
\end{proof}



\section{Experimental Results}\label{sec:exp}
There are two facts that make an experimental evaluation of the technique worthwhile: (1) Our reduction sets are necessarily {\em incomplete}. There may exist a general semantic reduction of the program (in the sense of Definition \ref{def:sr}) with a simple proof, but this reduction may not belong to the set of S-reductions or C-reductions defined in this paper. Therefore, an experimental evaluation to see how well the incomplete reductions fare in practice is essential. (2) The worst case time complexity of our algorithm is exponential, and therefore, it is important to know if an implementation of this algorithm can handle realistic examples.

\subsection{Implementation}

We have implemented our approach in a tool called \tool written in Haskell. \tool accepts a program written in a simple imperative language. The input language supports integers, booleans, arrays, uninterpreted functions, deterministic and non-deterministic branches and loops, parallel composition, assume statements, and assignment statements. The desired safety property is encoded in the input program itself in the form of \emph{assume} statements, and \tool attempts to prove the program safe.

\tool implements all of the optimizations of Section \ref{sec:optimization}. Note that since the algorithm as of Section \ref{sec:optimization} computes a fixpoint of a function over the product state space of the program and proof DFAs, no tree automata are ever explicitly constructed during an execution of the tool.

\subsubsection*{\bfseries SMT Solvers}

\tool supports a variety of background solvers. Only a few solvers support interpolation, but \tool can use different solvers for interpolation and proof generalization. For interpolation, \tool supports Z3, MathSAT, and SMTInterpol. For proof generalization, \tool additionally supports Yices and CVC4.
The main reason for multiple solver support is the general fragility of the interpolation tools. For example, MathSAT does well on some of our arithmetic benchmarks, but bugs out easily with the array benchmarks, while SMTInterpol does better with array interpolants. On the other hand, MathSAT performs better when it works.


\subsubsection*{\bfseries Counterexamples}

The set of counterexamples that provide the convergence guarantee of Theorem \ref{thm:stp} are often too large to be practically useful. It turns out that the algorithm converges in the strong majority of the cases if one selects only one counterexample from this set to move forward. The algorithm may take a few more refinement rounds to converge this way, but each round executes much faster and the overall time for verification ends up being substantially lower.
The choice of  counterexample can have a substantial impact on the total verification time. One can imagine many heuristics for this selection. We use two specifically for the evaluation in this section: one that picks a (mostly) sequentialized trace from all available traces (S), and another one which picks a mostly interleaved (I) counterexample; that is, it uses the counterexample that is going through the steps of different threads in a round-robin manner.

Recall the example in Figure \ref{fig:me1}. For verification with contextual (semi) reductions, the time under the (I) counterexample selection criterion is three times slower than the one under (S), since the {\em good} reduction is the sequential reduction. Big gaps like this one (in either direction) are observed in most benchmarks.

\subsection{Evaluation}

The target programs for our approach are those where a proof for the entire program is out of the reach of current automated verification tools due to the expressivity of the language of required interpolants.  For this reason, \tool cannot be compared against existing tools, as the premise is that they should fail on the majority of these benchmarks.

Given the same proof, checking it against an infinite set of reductions, in contrast to a single program in classic verification, is bound to be (theoretically) more expensive. Therefore, when not needed, reductions can cause a potentially large overhead on verification time. The exception is the cases where they are not strictly needed (from the theoretical point of view) but using them leads to a much smaller proof (in terms of the total number of assertions). In these scenarios, the smaller proof can offset the overhead of proof checking against reductions and lead to a better overall verification time.

\subsubsection*{\bfseries Benchmarks}

We have a diverse set of benchmarks in Table \ref{tab:res} which includes programs that require the reductions presented in this paper to be verified by an automated prover.
In other words, they are theoretically beyond the reach of the automated provers.
The reason for this is that a Floyd-Hoare style proof for the entire program (i.e. unreduced) will require a rich language of assertions that reason about  (1)  unbounded message buffers, (2) non-linear constraints, or (3) quantified facts about arrays, or a combination of these features. The benchmarks are arranged in Table \ref{tab:res} based on the complexity of these required assertions. \tool manages to prove these benchmarks correct by discovering a reduction for which the base theories of Linear Integer Arithmetic (LIA), Uninterpreted Functions (UF), and theory of arrays (unquantified).


Our second set of benchmarks, reported in Table \ref{tab:res2}, includes programs for which the S/C-reductions presented in this paper are not strictly required. They either require no reductions at all or the sleep-set reductions (from \cite{cav19}) are sufficient for proving them correct automatically. We use this second set of benchmarks to highlight the fact that the rich set of reductions presented in this paper can be of practical importance even if not theoretically required.

\begin{wraptable}{r}{5.7cm}\vspace{-10pt}
{\small
\begin{tabular}{|c|c|c|c|}\hline
  \multirow{2}{*}{Benchmark} & \multicolumn{3}{c|}{Reduction Style} \\ \cline{2-4}
 & S + C & C & S \\\hline
    \multicolumn{4}{|c|}{Unbounded Buffers} \\\hline
    channel-sum & 1.8 & {\bf 0.8} & {\sf TO} \\
    horseshoe & {\bf 45.2} & 45.5 & {\sf TO} \\
    prod-cons & 7.5 & {\bf 3.4} & {\sf TO} \\
    prod-cons-3 & 138.9 & {\bf 26.8} & {\sf TO} \\
    prod-cons-eq & {\bf 3.4} & 4.3 & {\sf TO} \\
    queue-add-2 & 6.2 & {\bf 5.9} & {\sf TO} \\
    send-receive & 5.7 & {\bf 4.9} & {\sf TO} \\
    send-receive-alt & 1.1& {\bf 0.5}& {\sf TO} \\
    simple-queue & 0.3& {\bf 0.03}& {\sf TO} \\
    queue-add-3 & {\bf 214.4}& {\sf TO} & {\sf TO} \\
    \hline
    \multicolumn{4}{|c|}{Nonlinear} \\\hline
    Figure \ref{fig:me2} & {\sf TO} & {\bf 0.3}& {\sf TO} \\
    mult-4 & {\sf TO} & {\bf 25.5}& {\sf TO} \\
    mult-equiv & 197.9& {\sf TO} & {\bf 194} \\
    counter-fun & {\bf 0.8}& {\sf TO} & {\sf TO} \\
    \hline
    \multicolumn{4}{|c|}{Arrays}\\\hline
    simple-array-sum & {\bf 52.4}& 97.4& {\sf TO} \\
    three-array-min & {\bf 39.1}& 74.2& {\sf TO} \\
    three-array-sum & {\bf 28.1}& 58.4& {\sf TO} \\
    three-array-max & {\sf TO} & {\bf 34.6}& {\sf TO} \\
    \hline
    \multicolumn{4}{|c|}{Unbounded Buffers + Nonlinear} \\\hline
    buffer-mult & 131.5& 212.4& {\sf TO} \\
    buffer-series & 62.9& 216.7& {\sf TO} \\
    buffer-series-array & 97.9& 292.2& {\sf TO} \\
    queue-add-2-nl & 14.4& 15.7& {\sf TO} \\
    queue-add-3-nl & 344& 314& {\sf TO} \\
    \hline
    \multicolumn{4}{|c|}{Queues + Arrays} \\\hline
    dec-subseq-array & 4.6& 7.3& {\sf TO} \\
    inc-subseq-array & 4.5& 6.7& {\sf TO} \\
    \hline
\end{tabular}} \vspace{2pt}
\caption{Experimental Results. Times are in seconds. Best times are in boldface. {\sf TO} indicate a timeout (set at 20mins). {\sc Weaver} is the tool from \cite{cav19}.}\label{tab:res}\vspace{-38pt}
\end{wraptable}

Unbounded buffers are modelled in these benchmarks using uninterpreted functions. More
precisely, a buffer is modelled using a triple $\left<f, i_0, i_1\right> \in (\mathbb{Z} \to \mathbb{Z}) \times \mathbb{Z} \times \mathbb{Z}$ where $f(i)$ denotes the $i$th element in the buffer, $i_0$ points to the first element in the buffer, and $i_1$ points to the last.

\subsubsection*{\bfseries Results}

We ran \tool on the benchmarks on a Dell Optiplex 3050 with an Intel(R) Core(TM) i7-7700 CPU (4 cores, 2 threads per core) and 32GB of RAM, running 64-bit Ubuntu 18.04.
The results are reported in Tables \ref{tab:res} and \ref{tab:res2}.
\tool has an option to turn semi-commutativity on and off, and we used it to measure the impact of it alone, and also when added to contextual commutativity relations. Note that C-reductions (of Definition \ref{def:conc}) are by default defined based on contextual semi-commutative relations, and therefore, they correspond to the  ``S + C'' option in Table \ref{tab:res}. The ``C'' column corresponds to C-reductions without semi-commutativity.

The ``{\sc None}'' column in the table corresponds to our implementation of \cite{HeizmannHP09} which does not perform reductions or any optimizations specific to handling concurrent programs. Therefore, it can be considered as a baseline algorithm. Our benchmarks are not very large programs. It is unlikely that a proof is not found by this baseline algorithm due to known intractability issues of concurrent program verification (i.e. state-space explosion). There are two reasons for failure: (1) the proof for the program is beyond capabilities of state-of-the-art SMT solvers (for interpolation and verification-condition checking), and (2) the algorithm falls into the well-known divergent behaviour of automated verification where sufficiently strong loop invariants are not produced.

All benchmarks in Table \ref{tab:res} fall in category (1). From Table \ref{tab:res2}, the benchmarks under Arrays also fall in category (1). Without S-reductions, C-reductions, or sleep-set reductions of \cite{cav19}, there would be a need for universal quantification over array elements. The rest of benchmarks in Table \ref{tab:res2}, for which the ``{\sc None}'' algorithm timeouts, fall under category (2). In these instances,  \tool succeeds with reductions because it gets lucky with the counterexamples of the reduction-based method and sidesteps the divergence issues. The example in Figure \ref{fig:me1} is one of these examples. If we modify the precondition to add the assertion {\tt \{ M = N \}} and change the postcondition to {\tt \{ y =  N - M \}}, then interpolation-based proofs do not diverge. This example is listed in Table \ref{tab:res2} as ``Figure \ref{fig:me1} (alt)''.

The results clearly demonstrate that C-reductions are very powerful in producing proofs in the majority of cases. There are cases that S-reductions alone make proving a program possible and there 
 \begin{wraptable}{r}{8.8cm}\vspace{-10pt}
{\small
  \begin{tabular}{|c|c|c|c|c|c|c|}\hline
    \multirow{2}{*}{Benchmark} & \multicolumn{5}{c|}{Reduction Style} \\ \cline{2-6}
    & S + C & C & S &{\sc Weaver} & {\sc None} \\\hline
    \multicolumn{6}{|c|}{Arrays + Nonlinear} \\\hline
    dot-product-array & 62.2& 105& {\bf 50.8} & 54.6& {\sf TO} \\
    \hline
    \multicolumn{6}{|c|}{Arrays} \\\hline
    max-array-hom & 1644 & 906& {\bf 756}  & 1483 & {\sf TO} \\
    max-array & 232& 306& {\bf 36.2}  & 446& {\sf TO} \\
    min-array-hom & 830& 1366 & {\bf 578}  & {\bf 578} & {\sf TO} \\
    min-array & 151.5& 180& {\bf 51.9} & 56.9& {\sf TO} \\
    sum-array-hom & 116& 167& {\bf 115} & 123& {\sf TO} \\
    sum-array & 64& 94.5& {\bf 46.5}  & 50.9& {\sf TO} \\
    parray-copy & 311& 407& {\bf 218} &  232& {\sf TO} \\
    mts-array & {\sf TO} & {\sf TO} & {\bf 1176} & 1190 & {\sf TO} \\
    sorted & {\sf TO} & {\sf TO} & {\bf 819} & {\sf TO} & {\sf TO} \\
    \hline
    \multicolumn{6}{|c|}{Unbounded Buffers} \\\hline
    commit-1 & 3.2& 5.3& {\bf 1.7} &  1.7& 1.9\\
    commit-2 & 15.9& 38.4& {\bf 6.4} &  14.4& 31.1\\
    two-queue & 8.1& {\bf 2.6} & 15.1&  15.2& 57.7\\
    \hline
    \multicolumn{6}{|c|}{Standard Language of Assertions} \\\hline
    Figure \ref{fig:me1} & 0.21& {\bf 0.2} & {\sf TO} & {\sf TO} & {\sf TO} \\
    Figure \ref{fig:me1} (alt) & 1.4& 2.1& {\bf 1.2} &  3.0& 3.2\\
    counter-determinism & {\bf 5.8} & 10.5& {\sf TO} & {\sf TO} & {\sf TO} \\
    difference-det & 25.9& 25.1& {\bf 12.5} & {\sf TO} & {\sf TO} \\
    nonblocking-cntr & 1.3& {\bf 1.1} & {\sf TO} & {\sf TO} & {\sf TO} \\
    nonblocking-cntr-alt & {\bf 3.7} & 3.8& 19.3&  28.1& {\sf TO} \\
    min-le-max & 3.2& 2.7& 0.2& {\bf 0.1} & 0.4\\
    threaded-sum-2 & {\bf 3.1} & 3.4& 9.5& 10& 3.4\\
    threaded-sum-3 & 139& 90.7& {\bf 84.2} & {\sf TO} & {\sf TO} \\
    \hline
  \end{tabular}}
\caption{More experimental results. Times are in seconds. Best time for each benchmark appears in boldface. {\sf TO} indicate a timeout (set at 30mins). {\sc Weaver} is the tool from \cite{cav19}. {\sc None} is without any reductions \cite {HeizmannHP09}. }
\label{tab:res2}\vspace{-20pt}
\end{wraptable}
are cases that semi-commutativity substantially boosts contextual reductions. Theoretically, adding the option of semi-commutation specifications should only make the tool perform better. However, a change in the independence relation could result in a change in the counterexamples used in the refinement rounds, and in four of the benchmarks (from both tables), this change seems to be adversarial for the algorithm; in these cases, the contextual reductions without semi-commutativity manage to produce a proof, but adding in semi-commutativity causes timeouts.

Other than a few exceptions, the counterexample selection strategy (I) described before produces the fastest time over the benchmarks. With the solvers the results are mixed. The best times are split (near half-half) between MathSAT and SMTInterpol.

On average 13 refinement rounds are required to verify the benchmarks, with 43 being the maximum number. The average proof size is about 93 assertions and the largest proof includes 340 assertions. Of the benchmarks that take more than 5 seconds to verify, the majority are three-threaded programs. For these, on average, 81\% of the time is spent in proof construction, 11\% in proof checking, and 8\% in interpolation. For the six four-threaded benchmarks, the averages are different: 38\% of the time is spent in proof construction, 54\% in proof checking, and 8\% in interpolation. It is expected that as the number of threads increases, the cost of proof checking should dominate the total verification since it increases exponentially with the number threads.

\tool and all of our benchmarks are available at {\tt \url{https://github.com/weaver-verifier/weaver}}.

\subsection*{The Optimized Proof Checking Algorithm}

In Section \ref{sec:optimization}, we proposed a novel way of devising a faster proof checking algorithm. We evaluate this algorithm separately by comparing the times taken for proof checking a complete proof for each benchmark in the standard algorithm versus the optimized algorithm. The optimized algorithm performed 6.7x faster than the standard one in the best case, and 1.7x faster on average. In the worst case, the optimized algorithm's performance was the same as the standard algorithm (in exactly one case).  Note that this is an isolated evaluation of the algorithms for poofs that check. In the refinement loop, most proof checking tests fail, and using the optimized version produces better overall speedups than these reported numbers. Since the two algorithms may produce different counterexamples, one cannot compare the overall time of the entire verification process between the two proof checking algorithm choices. There is no guarantee that the speedups (or slowdowns) observed are not due to better (or worse) luck with the selection of counterexamples.


\section{Related Work}
The contributions of this paper relate to several topics, including automated concurrent program verification, relational verification, and program reductions. Each topic has a vast literature of related work. Here, we only explore connections to the most relevant work. Specifically, a large body of related work using reduction for the purpose of bug finding (in contrast to producing proofs) is not discussed  since the focus of this paper is on sound reductions for verification.

\subsection*{Reductions for Concurrent Program Verification}

Lipton's reduction \cite{Lipton75} has inspired several approaches to concurrent program verification \cite{ElmasQT09,civl,KraglQH18}, which fundamentally opt for inferring large atomic blocks of code (using various different techniques) to leverage mostly sequential reasoning for concurrent program verification. QED \cite{ElmasQT09} and CIVL \cite{civl} frameworks both use refinement-oriented approaches to proving concurrent programs correct. These {\em semi-automatic} systems use a combination of ideas to simplify proofs of concurrent programs. Specifically, yield predicates (location invariants) are similar to the contexts for commutativity in this paper.
CIVL\cite{civl} takes advantage of classic movers wherever applicable, so as not to have to rely too heavily on yield predicates. QED \cite{ElmasQT09} performs small rewrites in the concurrent program that have to be justified by potentially expensive reduction and invariant reasoning. Both systems are more broadly applicable since they deal with functions and subroutines which are not part of our program model.

In a different direction, program reductions (beyond atomicity specifications) have been used to simplify concurrent and distributed program proofs by eliminating the need to reason about unbounded message buffers.
In \cite{Genest07}, the theory of Mazurkiewicz traces is used to define a category of distributed systems, modelled as automata communicating through channels, which are {\em existentially bounded}. {\em Natural proofs} \cite{Desai14} and {\em pretend synchrony}\cite{psynch} (among many more) use the same fundamental idea to simplify reasoning about distributed systems. For the programs which are targets of these approaches, large atomic blocks are not a reduction of choice  since the aim of the reduction (i.e. program simplification) is to simplify the program from an asynchronous to almost synchronous.

Natural proofs \cite{Desai14} work for unbounded domains but boundedly many processes. {\em Pretend synchrony} \cite{psynch} provides an extension that works with unboundedly many processes by rewriting the program into an equivalent synchronous one. To make this possible, however, assumptions are made about loops (that there is no loop state) and round non-interference (no carried state between rounds). These are reasonable assumptions for distributed protocols but do not apply to concurrent message-passing programs. We also assume boundedly many processes, simply to be able to use finite state automata. Limited notions of context appear in some domain-specific reduction techniques. For example, in natural proofs \cite{Desai14}, it matters whether a buffer is empty or not. Contextual reductions, however, are more general than context specific to buffers.


We emphasize that all these techniques are incomplete, even for the particular domains for which they were designed. Contextual reductions are also incomplete. As already noted in \cite{cav19}, the problem of finding a reduction is as difficult as proving safety.

\subsection*{Partial Order Reduction}
Partial-order reduction (POR) \cite{Godefroid96,AbdullaAJS17,AbdullaAJS14} is a class of techniques that reduces the state space of search (for violation of a safety property) by removing redundant paths. POR techniques are concerned with finding a single (preferably minimal) reduction of (mostly) finite-state systems, and their primary application is in reachability/unreachability queries. We use the underlying ideas in POR in a non-standard way. The design of the LTAs that recognize S-reductions and C-reductions are informed by them.

Context has been incorporated into POR algorithms before. In \cite{KatzP92,GodefroidP93}, conditional dependence is used as a weakening of the independence relation to increase the potential for reduction. Conditional dependence adds a third component to the dependence relation, which is a (single) state. These techniques are exclusively applicable to finite-state systems. One can view one of the contributions of this paper as providing a way of lifting these ideas to infinite-state programs. In \cite{peep}, the notion of {\em guarded dependence} is introduced which extends the state to a predicate (i.e. a set of states). This is then used to perform POR in the context of bounded symbolic model checking of finite state systems.

A language-theoretic notion of context has been previously  studied in the context of models of concurrency \cite{gtrace}. Our language-theoretic definition can be viewed as a weakening of that notion of Generalized Mazurkiewicz Trace Languages, which have additional {\em consistency} and {\em coherence} conditions on relation $\mathcal{I}$. 

Partial order reduction has been combined with automated verification methods to tackle the large state space of multithreaded programs  \cite{por-impact, por-ta, WangCGY09}. In \cite{por-impact}, POR is combined with the classic {\sc IMPACT} algorithm to lift it to concurrent programs. In \cite{por-ta}, POR is applied to the concurrent control-flow automaton of the program to construct a reduced one, which is then used for proof construction/checking in a classic refinement algorithm in the style of \cite{HeizmannHP09}.
Contexts do not play (a significant) role in pruning mechanisms of either of the approaches presented in \cite{por-impact,por-ta}.



\subsection*{Relational and Hypersafety Verification}
Program reductions have been used in relational and hypersafety verification \cite{PnueliSS98,GoguenM82a,SabelfeldM03,SousaDVDG14,BartheCK11,SousaD16} where reductions are applied to product programs to obtain simple proofs of relational/hyper- properties. The important observation is that since the copies of the program in such product programs are completely disjoint, the statements fully commute for the purpose of constructing a reduction. The contributions of this paper become significant when one does not have such a trivial commutativity relation. For example, if the goal is to prove a relational/hyper property of a concurrent program, where beyond the top-level product, commutativity specifications within a copy become relevant. We have examples of this (for instance proving the determinism of a concurrent program) among our benchmarks in Section \ref{sec:exp}.  Neither of the approaches cited can handle concurrent programs.

The work in \cite{cav19} is the closest to ours in terms of methodology and in the fact that it handles concurrent programs. There, in the same style of refinement loop, the space of {\em non-contextual} reductions based on a {\em symmetric} dependence relation are explored for the purpose of verification of hypersafety properties of sequential and concurrent programs.  Programs proved correct in this paper that require reasoning about semi-commutativity and contextual commutativity are theoretically beyond the scope of the algorithm presented in \cite{cav19}.




Equivalence checking of looped programs has been explored as an instance of relational verification \cite{symdiff,Sharma13,ChurchillP0A19}. In \cite{Sharma13} loops are never unrolled and therefore the approach is limited to cases that a simple proof exists for unrolled loops. SymDiff \cite{symdiff} uses (unsound) unrolling of the loops for a fixed number of iterations for verification. And, finally and most recently, in \cite{ChurchillP0A19}, concrete executions are used as a guide to guess a good correspondence between (potentially unrolled) executions of the loops to push the frontier further. In our approach, arbitrary (unbounded) unrollings of loops are considered for constructing proofs through the reduction automaton. As long as the correct invariants are guessed through the interpolation method, the approach can succeed in finding the right correspondence and a proof for it if one exists.


\section{Conclusion and Future Work}
The notion of {\em context} for program reductions had not received much attention before this paper. C-reductions provide a solution to incorporate contextual reductions in the automated program verification tool. The preliminary experimental results (Section \ref{sec:exp}) are promising. Nontrivial examples, that previously could not be proved automatically, can be verified using \tool. There is, however, much more work left ahead to explore the full potential of program reductions for automated program verification.

First, in Section \ref{sec:optimization}, we presented algorithmic optimizations that ensure no additional complexity is incurred for contextual reductions over the simpler reductions of \cite{cav19}. The proof checking algorithm however still has a high complexity. This may be acceptable as a worst-case complexity for proof checking, but the construction of section \ref{sec:optimization} implicitly requires a construction of the program control flow automaton, whose size exponential on the number of threads. Consider the case of a very simple parallel programs where all threads are {\em disjoint} and the postcondition refers only to the variables in a single thread. \tool can handle proving this program for a small number of threads, but as the number of threads grows, \tool's verification time grows exponentially with it. It will be interesting to explore alternative ways of defining the reduction LTAs and/or the exploration algorithms to find solutions with better average case complexity when the number of threads grows but the verification task remains simple. For example, exploiting symmetry for replicated code is one possible avenue of investigation.

Second, it would be interesting to see how the idea of {\em abstraction} used by QED \cite{ElmasQT09} and CIVL \cite{civl} can be incorporated in the framework put forward by this paper to gain more powerful reductions. Briefly, a non-commuting statement can be abstracted in way that the new program still satisfies the property of interest and the new abstract statement commutes against more statements than the old concrete one. These abstractions are suggested manually in \cite{ElmasQT09,civl} and it will be interesting to investigate if the same insight can be inferred automatically.

Another limitation of the current approach is that it works for a fixed number of threads. It would be interesting to explore if {\em predicate automata} \cite{FarzanKP15} or {\em nominal automata} \cite{BojanczykKL14} can be used to formulate reductions for  parameterized concurrent programs.

\newpage
\bibliographystyle{plainnat}
\bibliography{popl20}

\newpage
\appendix

\end{document}